\documentclass[aps,pra,twocolumn,superscriptaddress,longbibliography]{revtex4-1}
\usepackage{times}
\usepackage{graphicx, amsmath, verbatim, dsfont, amsfonts, color}
\usepackage{bbm, bm, latexsym, amssymb}
\usepackage{graphicx}% Include figure files
\usepackage{epstopdf}
\usepackage{dcolumn}% Align table columns on decimal point
\usepackage{bm}% bold math
\usepackage{gensymb}
\usepackage{upgreek}
\usepackage{hyperref}
\usepackage{graphicx}% Include figure files
\usepackage{epstopdf}
\usepackage{dcolumn}% Align table columns on decimal point
\usepackage{bm}% bold math
\usepackage{gensymb}
\usepackage{upgreek}
\usepackage{booktabs}
\usepackage{hyperref}

\usepackage{amssymb,mathtools}
\usepackage{color}
\usepackage{float}
\begin{document}
\author{Tomasz~Dietl}
\email{dietl@MagTop.ifpan.edu.pl}
\affiliation{International Research Centre MagTop, Institute of Physics, Polish Academy of Sciences, Aleja Lotnikow 32/46, PL-02668 Warsaw, Poland}
\affiliation{WPI Advanced Institute for Materials Research, Tohoku University, 2-1-1 Katahira, Aoba-ku, Sendai 980-8577, Japan}

\title{Quantitative theory of backscattering in topological HgTe and (Hg,Mn)Te quantum wells:\\
 acceptor states, Kondo effect, precessional dephasing, and bound magnetic polaron}

\begin{abstract}
We present the theory and numerical evaluations of the backscattering rate determined by acceptor holes
or Mn spins in HgTe and (Hg,Mn)Te quantum wells in the quantum spin Hall regime. The role
of anisotropic $s$--$p$ and $sp$--$d$ exchange interactions, Kondo coupling, Luttinger liquid effects, precessional dephasing,
and bound magnetic polarons is quantified. The determined magnitude and temperature dependence
of conductance are in accord with experimental
results for HgTe and (Hg,Mn)Te quantum wells.
\end{abstract}
%\clearpage
\maketitle

\section{Introduction}
The experimental discovery of the quantum spin Hall effect (QSHE) in HgTe quantum wells (QWs) \cite{Konig:2007_S} provided experimental support for the seminal theoretical predictions about the topological protection \cite{Kane:2005_PRL,Bernevig:2006_S} and opened prospects for electron transport without scattering backward, the capability of interest for energy-efficient and decoherence-free classical and quantum devices. However, in studied two-dimensional (2D) topological systems, such as aforementioned HgTe QWs \cite{Konig:2007_S,Roth:2009_S} and 1T'-WTe$_2$ 2D monolayers \cite{Fei:2017_NP,Wu:2018_S}, the two-terminal conductance approaches the quantized value $2e^2/h$ only in mesoscopic samples, shorter than 10 and 0.1\,$\mu$m, respectively, without much improvement on cooling below 50\,mK.

Understandably, the origin of processes accounting for the unanticipatedly short topological protection length has attracted a considerable attention. In 2D topological materials, electrons reside in counter-propagating Kramers degenerate helical states adjacent to sample edges. Furthermore, the direction of electron momentum determines the spin orientation. Therefore, mechanisms leading to electron backscattering  must contain ingredients that break both time-reversal and spin-rotation symmetry.  As reviewed recently \cite{Hsu:2021_SST,Yevtushenko:2022_NJP},  previous theories considering backscattering in the linear response regime can be costed into three categories. The first of them contains approaches associating those ingredients to the presence of electron-electron interactions within the helical channels \cite{Strom:2010_PRL,Crepin:2012_PRB,Lezmy:2012_PRB,Pikulin:2014_PRL,Wang:2017_PRL,Novelli:2019_PRL}. The second class of models considers the effects of external spins coming from extrinsic defects \cite{Maciejko:2009_PRL,Altshuler:2013_PRL,Vayrynen:2014_PRB}, magnetic impurities \cite{Hattori:2011_JPSJ,Tanaka:2011_PRL,Cheianov:2013_PRL,Kimme:2016_PRB,Kurilovich:2019_PRB,Yevtushenko:2022_NJP} or nuclear spins \cite{Lunde:2013_PRB,Hsu:2017_PRB}. Finally, the influence of random magnetic fluxes has been examined \cite{Delplace:2012_PRL}. It can be argued that the abundance of theoretical proposals reflects  difficulties in assessing magnitudes of material parameters entering into particular models, precluding a conclusive comparison of the theory to experimental results.

This paper, supporting and extending the companion report \cite{Dietl:2023_PRL}, presents theory and provides numerical evaluations of conductance in the regime of the quantum spin Hall effect (QSHE) in HgTe \cite{Konig:2007_S} and (Hg,Mn)Te quantum wells \cite{Shamim:2020_SA}, considering holes localized on acceptors as a source of backscattering. We have quantitatively determined acceptor energies, the Coulomb gap, exchange coupling between edge electrons and acceptor holes, Kondo temperatures for both acceptor holes and Mn spins, spin-flip and backscattering rates in  strong and weak coupling regimes, and effects of bound magnetic polarons. In addition to the Kondo effect discussed already in the context of backscattering \cite{Maciejko:2009_PRL,Altshuler:2013_PRL}, we incorporate into our approach spin-nonconserving anisotropic components of the exchange coupling, allowing for backscattering even in the spin-momentum locking case \cite{Tanaka:2011_PRL,Altshuler:2013_PRL,Lunde:2013_PRB,Kimme:2016_PRB,Yevtushenko:2022_NJP}.

Our theoretical and numerical results substantiate the main conclusion of the companion paper \cite{Dietl:2023_PRL}  that the value of the exchange interaction between edge electrons and acceptor holes is large enough to drive the system to the strong coupling regime of the Kondo effect, where the spin-flip scattering rate attains the unitary limit. Together with the demonstrated sizable exchange anisotropy, the unitary scattering rate explains the observed topological protection length in HgTe QWs in the Ohmic conductivity regime \cite{Konig:2007_S,Bendias:2018_NL,Lunczer:2019_PRL} and the corresponding temperature dependence of conductance, if the influence of Luttinger liquid effects upon the anisotropic exchange are taken into account \cite{Vayrynen:2016_PRB}. A key aspect of our approach is the observation that the acceptor concentration is known experimentally, as it determines hole density in undoped QWs and the gate voltage range separating the electron and hole transport in a given QW. In this way our theory does not involve any fitting parameters. Accordingly, the quantitative agreement with experimental data appears meaningful.

In contrast to occupied acceptor dopants, our results indicate that {\em no} Kondo effect is expected for impurities with open $d$ or $f$ shells in dilute magnetic semiconductors (DMSs), at least for typical values of antiferromagnetic $sp$--$df$ exchange integrals. Furthermore, for magnetic ions with orbital momentum $L =0$, such as Mn$^{2+}$, anisotropic components of the exchange coupling to edge states vanish, suggesting a minor role of Mn spins in backscattering. Nevertheless, we suggest that backscattering by a dense bath of interacting Mn spins can operate,  as for the motionally-narrowed precessional spin-dephasing process, the constraint of local spin momentum conservation is relaxed. At the same time, we show that the formation of bound magnetic polarons by acceptor holes reduces the role of the Kondo effect, the observation elucidating the origin of the recovery of quantized conductance at low temperatures in (Hg,Mn)Te QWs \cite{Shamim:2021_NC}.
\section{Theoretical results}
\subsection{Quantum-well band states}
\label{sec:bands}
A starting point of our theory is the eight-band Kohn--Luttinger Hamiltonian with boundary conditions previously used
to obtain the subband structure in HgTe QWs \cite{Novik:2005_PRB} and the four--band model developed for determining,
in the axial approximation, acceptor levels in GaAs QWs \cite{Fraizzoli:1991_PRB}. Within that approximation, discussed in Sec.\,\ref{sec:bands}, and in the absence of the impurity potential,
the eight relevant electron wave functions $\psi_i$ can be taken in the form,
\begin{equation}
\psi_i^{(m_i)}(\rho,\varphi,z) = h_i^{(m_i)}(\rho,\varphi)f_i(z)u_i,
\end{equation}
where $m_i$ is the integer orbital quantum number corresponding to the $z$-component of the orbital momentum $l_z$ of the envelope function $h_i$;
$\rho$, $\varphi$, and $z$ denote the electron position in the cylindrical coordinates with $z$ along the growth direction;
$u_i \equiv u_{j,jz}$ is the set of the Kohn--Luttinger amplitudes for particular angular momenta $j$ and its $z$-component $j_z$: $s_{1/2,1/2}$, $s_{1/2,-1/2}$, $p_{3/2,3/2}$, $p_{3/2,1/2}$, $p_{3/2,-1/2}$, $p_{3/2,-3/2}$, $p_{1/2,1/2}$,
and $p_{1/2,-1/2}$ \cite{Novik:2005_PRB}, whereas the envelope functions $h_i^{(m_i)}(\rho,\phi)$ and $f_i(z)$ are given by,
\begin{equation}
h_i^{(m_i)}(\rho,\varphi) = N_{\text{o}}i^{|m_i|}\exp(im_i\varphi)J_{|m_i|}(k\rho),
\label{eq:Bessel}
\end{equation}
where $N_{\text{o}}$ is a normalization factor. The Hamiltonian in question can be diagonalized by taking
 $m_i = m$ for $i = 1$, 4, and 7; $m_i = m+1$ for $i = 2$, 5, and 8; $m_i = m-1$ for $i = 3$,
and $m_i = m + 2$ for $i=6$, where $m$ is an integer and  $J_{|m_i|}(k\rho)$ is the Bessel function; $k$ is a module of the in-plane wavevector, and
\begin{equation}
f_i(z) = (1/\sqrt{L_z})\sum_{n=-n_{\text{max}}}^{n_{\text{max}}}c^{(i)}_n\exp(2\pi inz/L_z),
\label{eq:f_z}
\end{equation}
where $L_z$ is the total thickness of the structure, including the two Cd$_{0.7}$Hg$_{0.3}$Te barriers and HgTe QW, assumed here as
$L_z = 60$\,nm\,+\,$d_{\text{QW}}$; the expansion coefficients $c^{(i)}_n$ are to be determined by a diagonalization procedure;
$n_{\text{max}} = 50$ ensures an appropriate numerical convergence, as discussed in Sec.\,\ref{sec:acceptors}. As seen by inspection, each of the eight components $\psi_i^{(m_i)}(\rho,\varphi,z)$ and, hence, the total wave function corresponds to the same value of the total angular momentum $F_z = l_z + j_z = m+1/2$, meaning that the operator $F_z$ commutes with the eight-bands' axial Hamiltonian.

We adopt the identical values of low-temperature ${\vec{k}}\cdot{\vec{p}}$ parameters, as in Ref.\,\cite{Novik:2005_PRB}, except for the Luttinger $\kappa$ that we assume to be the same in
the barriers and QW. In this way, no Rashba-like splitting occurs in symmetric QWs. Computations have been performed assuming the absence of strain. Its role is discussed in Sec.\,\ref{sec:strain}.

\begin{figure}[tb]
	%\centering
%\hspace*{-1cm}
	\includegraphics[width =0.95\columnwidth]{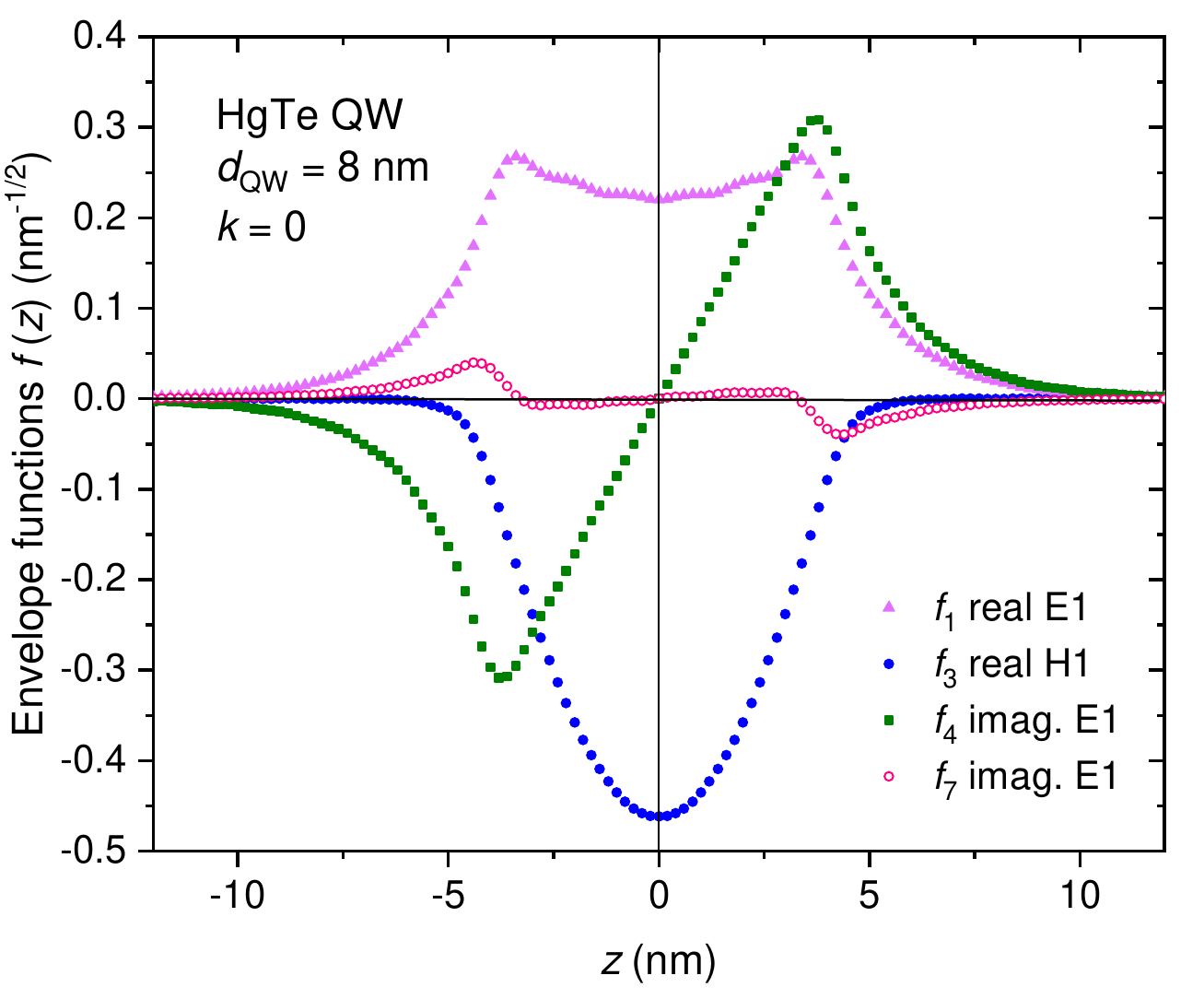}
	\caption{Envelope functions $f_1(z)$, $f_3(z)$, $f_4(z)$, and $f_7(z)$  at $k=0$ for E1 and H1 subbands in HgTe quatum well of 8\,nm-width. The envelopes $f_3$ and $f_6 = f_3$ correspond to the Kramer's pair for the H1 subband, whereas $f_1$, $f_4$, $f_7$  and $f_2 = -f_1$, $f_5 = -f_4$, $f_8 = f_7$ for the E1 subband with a relative weight 53.6, 45.9, and 0.5\%, respectively. As seen, the values of $f_j$ are either real and symmetric or imaginary and antisymmetric in respect to the QW center at $z =0$.}
\label{fig:envelope}
\end{figure}

Figures 1(a,b) in the companion paper \cite{Dietl:2023_PRL} present QW subband dispersions $E(k)$ for $d_{\text{QW}} = 6$ and 8\,nm.  The magnitudes of $f_i(z)$  for $d_{\text{QW}} = 8$\,nm and $k =0$ are shown in Fig.~\ref{fig:envelope}. A character of their inversion symmetry ($z \rightarrow -z$) plays an essential role in the anisotropy of the electron-hole exchange interaction, as discussed in Sec.\,\ref{sec:exchange}.

 \subsection{Axial approximation}
 \label{sec:axial}
\begin{figure}[tb]
	%\centering
%\hspace*{-1cm}
\hspace*{-0.5cm}
\hbox to \hsize{
	\includegraphics[scale=0.4, angle=0]{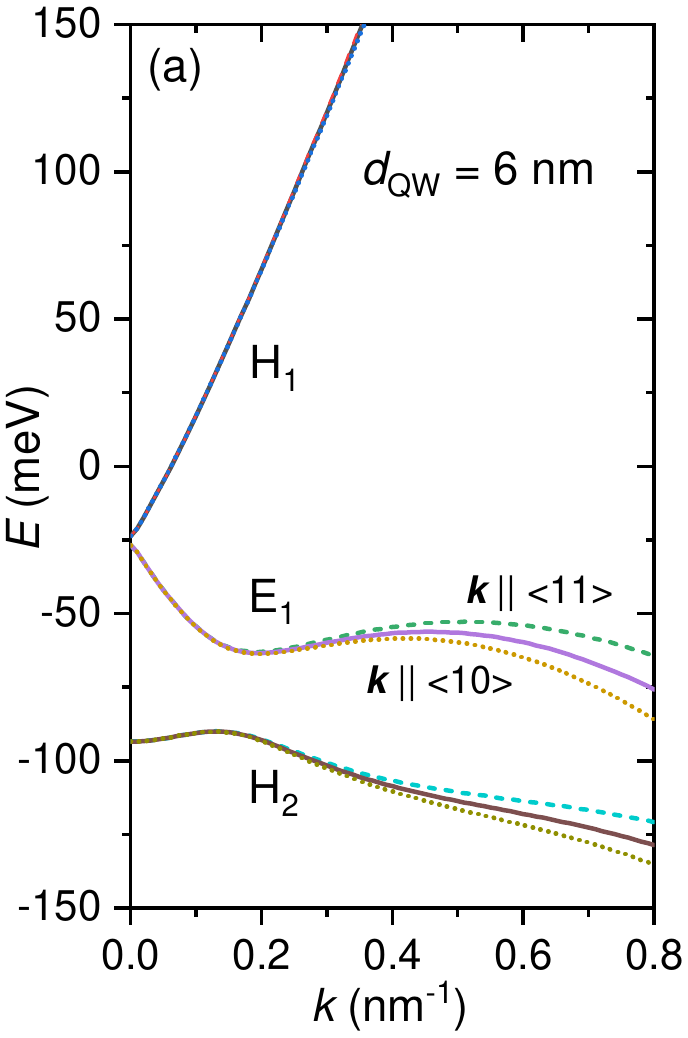}
\includegraphics[scale=0.4, angle=0]{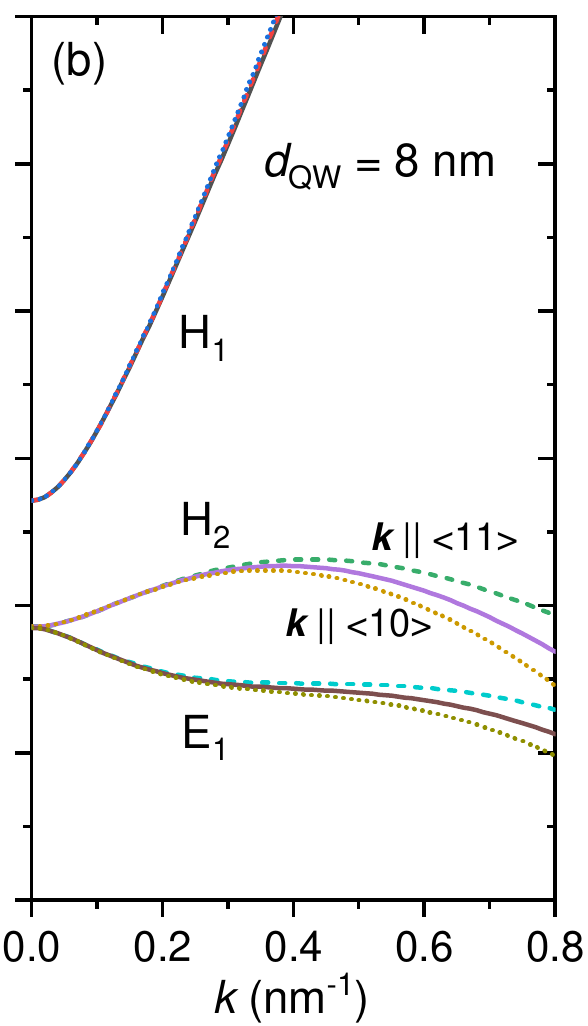}}
	\caption{ Verification of axial approximation. Subband dispersions for a HgTe QW of the thickness $d_{\text{QW}} =6$\,nm (a) and $8$\,nm (b) computed for
${\vec{k}}\| \langle11\rangle$ (dashed lines),  ${\vec{k}}\| \langle10\rangle$ (dotted lines) and within the axial approximation (solid lines). Character of particular subbands at $k = 0$ is also marked.}
\label{fig:axial}
\end{figure}

 Figure \ref{fig:axial} shows  subband dispersions $E({\vec{k}})$ computed without and with the axial approximation for HgTe QW thicknesses 6 and 8\, nm.
 A slight overestimation of the indirect gap by the axial approximation is visible for $d_{QW} =8$\,nm. Since, however, the energy differences between ${\vec{k}}\| \langle10\rangle$ and ${\vec{k}}\| \langle11\rangle$ near valence band top are significantly smaller than the acceptor binding energies in Fig.\,1(c) of the companion paper \cite{Dietl:2023_PRL}, the axial approximation holds.

 \subsection{Strain effects}
\label{sec:strain}
\begin{figure}[tb]
	%\centering
\hspace*{-0.5cm}
\hbox to \hsize{
	\includegraphics[scale=0.5, angle=0]{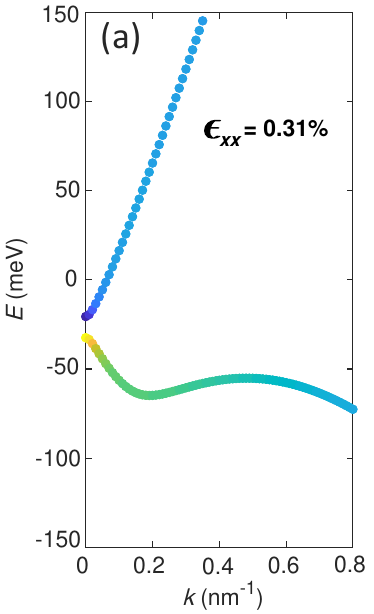}
\includegraphics[scale=0.5, angle=0]{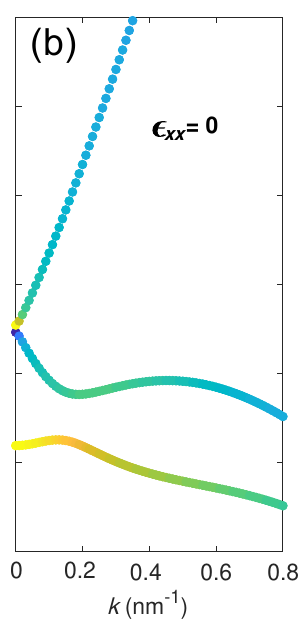}
\includegraphics[scale=0.5, angle=0]{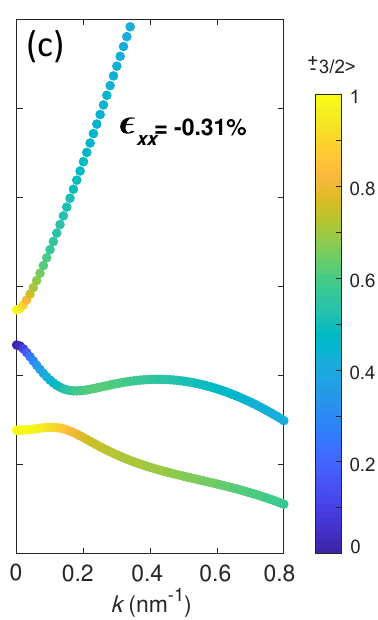}}
	\caption{Effects of biaxial strain on band dispersion in HgTe QWs. (a) $\epsilon_{xx} = 0.31$\% (CdTe substrate); (b) $\epsilon_{xx} = 0$; (c) $\epsilon_{xx} = -0.31$\%. Colors describe the participation of the $p_{3/2,\pm3/2}$ Kohn-Luttinger amplitude in the electronic wave function; HgTe QW thickness $d_{\text{QW}} =6$\,nm.}
\label{fig:strain}
\end{figure}

We use a conventional sign definition of biaxial epitaxial strain \cite{Bir:1974_B,Reilly:1989_SST,Dietl:2014_RMP}, i.e., a positive value of $\epsilon_{xx}=\epsilon_{yy}$ corresponds to QW under the tensile strain. In particular, in the case of HgTe QW,
$\epsilon_{xx} = 0.31$\% corresponds to strain for a CdTe substrate, whereas a Cd$_{1-x}$Zn$_{x}$Te substrate generates a compressive strain \cite{Leubner:2016_PRL}, $\epsilon_{xx} <0$. The evolution of the QW band structure with biaxial strain is shown in Fig.\,\ref{fig:strain} for HgTe QW of the  thickness $d_{\text{QW}} =6$\,nm and $\epsilon_{xx} = 0.31$\%, 0, and $-0.31$\%.
The strain, and also temperature, shift the QW thickness corresponding to the topological phase transition ($E_{\text{g}}=0$), but the overall band picture remains intact, in agreement with previous conclusions \cite{Novik:2005_PRB,Leubner:2016_PRL,Kadykov:2018_PRL}.

\subsection{Determination of acceptor level energies}
 \label{sec:acceptors}
 To determine energies of levels brought about by charge dopants,
 we supplement the Hamiltonian by the Coulomb potential,
 \begin{equation}
 V(r) = -Ze^2/4\pi\epsilon_0\epsilon r,
 \end{equation}
 and by the potential of image charges in the barriers \cite{Fraizzoli:1991_PRB}, for which the dielectric constant is
% \begin{equation}
 $\epsilon_{\text{b}} = (1-x_{\text{b}})\epsilon + x_{\text{b}}\epsilon_{\text{e}}$,
 %\end{equation}
 where in our case $\epsilon = 19.5$, $\epsilon_{\text{e}} = 10.5$, and $x_{\text{b}} =0.7$. We neglect central cell corrections and the image charge in the gate metal, which is typically more than 100\,nm apart.

Furthermore, we replace the Bessel function $J_{|m_j|}(k\rho)$ in Eq.\,\ref{eq:Bessel} by
\begin{equation}
J_{|m_i|}(k\rho) \rightarrow \rho^{|m_i|}\sum_{l=1}^{l_{\text{max}}}d^{(i)}_l\exp(-\rho/a_l),
\label{eq:exponential}
\end{equation}
where the coefficients $d^{(i)}_l$ are to be determined by the diagonalization procedure for a given set of ${a_l}$ values.
We take ${a_l}$ as a geometrical series with a common ratio of $r = 1.5$ and,  for a typical number of exponential functions $l_{\text{max}} =15$, the starting value of $a_1 =0.5$\,nm. Since, the exponential
functions with  real exponents are not orthogonal, a generalized eigenvalue solver has been
employed to obtain electronic energies. At the same time, the participation number serves to evaluate an effective in-plane localization radius $a^*$.

\begin{figure}[tb]
	%\centering
%\hspace*{-2.3cm}
	\includegraphics[width =\columnwidth, angle=0]{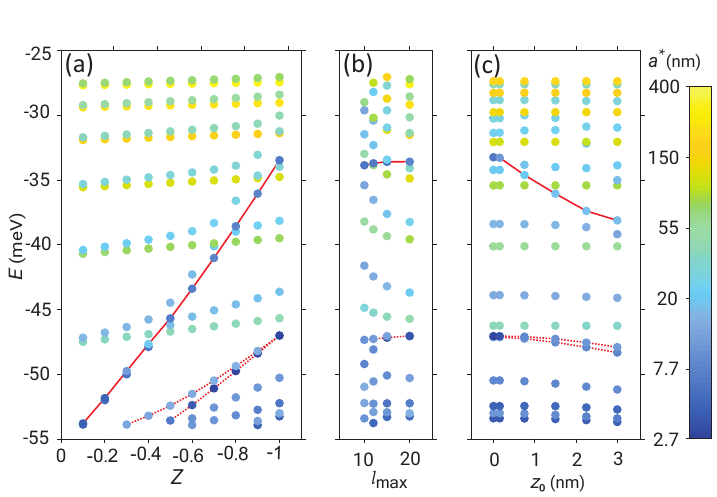}
	\caption{Example of the identification of acceptor levels $E_{3/2}$ $(m = 1)$ degenerate with the QW band states. (a) Energy levels $E$ as a function of the center charge $Z$. (b) $E$ as a function of the number of exponential functions $l_{\text{max}}$ in Eq.\,\ref{eq:exponential} and for $n_{\text{max}} =25$ (Eq.\,\ref{eq:f_z}). (c) $E$ as a function of the distance of acceptor to QW center $z_0$ for $l_{\text{max}} = 15$ and $n_{\text{max}} =50$. Color scale depicts the magnitude of an effective in-plane Bohr radius $a^*$ in the logarithmic scale. Red solid lines show the ground state $E_{3/2}$ acceptor level; dotted lines represent examples of acceptor excited states. HgTe QW thickness $d_{\text{QW}} =6$\,nm; no strain.}
\label{fig:levels}
\end{figure}

The presence
of the impurity potential breaks the degeneracy of the states with respect to the quantum number $m$.
However, due to time--reversal symmetry, the impurity levels remain at least
doubly degenerate. The use of exponential functions (Eq.\,\ref{eq:exponential}) is suitable for determining
the localized levels but not for oscillating extended states. Accordingly, the values of band energies do not converge with increasing $l_{\text{max}}$.

In our case, identifying acceptor levels that overlap with the
continuum of band states appears difficult \cite{Buczko:1992_PRB,Zholudev:2020_S}. Figure~\ref{fig:levels} depicts level energies determined for $m = 1$ in a HgTe QW with $d_{\text{QW}} = 6$\,nm. As shown, the magnitude of $a^*$, and the evolution of the level energies with $Z$ [Fig.\,\label{fig:test}(a)],  $l_{\text{max}}$  [Fig.\,\ref{fig:levels}(b)], and with the acceptor
location of the QW center [Fig.\,\ref{fig:levels}(c)] unambiguously tell the band and resonant impurity states. According to the result presented in the companion paper \cite{Dietl:2023_PRL}, the ground state corresponds to $m=0$ or $m =-1$, and is predominately built of the $p_{3/2,\pm1/2}$ and $s_{1/2,\pm1/2}$ Kohn-Luttinger amplitudes. Its energy is referred to as $E_{1/2}$.

\subsection{Coulomb gap}
\label{sec:gap}
The Efros-Shklovskii Coulomb gap \cite{Shklovskii:1984_B} of the width $E_{\text{C}}$ reaching 10\,meV was observed in a single-layer of 1T'-WTe$_2$ with different coverage of surface by potassium \cite{Song:2018_NC}. Furthermore, numerical simulations revealed the presence of the Coulomb gap for resonant donor states in HgSe:Fe \cite{Wilamowski:1990_SSC}.

In terms of 2D DOS of the acceptor band taken as  $\nu_2 =N_{\text{a}}/E_{\text{b}}$,  where $E_{\text{b}}$ is the acceptor bandwidth \cite{Shklovskii:1984_B},
\begin{equation}
E_{\text{C}} = 2\nu_2e^4/(4\pi\epsilon_0\epsilon)^2.
\end{equation}
In the case of QWs, a non-zero $E_{\text{b}}$ originates from a dependence of the binding energy on the distance between the impurity and QW center. Figure 1(c) in the companion paper \cite{Dietl:2023_PRL} shows this effect for acceptors in HgTe QWs, and demonstrates that the bandwidth extends down to a side maximum of the valence band characterized by a heavy mass. For a HgTe QW with $d_{\text{QW}} =8$\,nm, where $E_{\text{b}} \simeq 20$\,meV and the QW dielectric constant $\epsilon = 19.5$,   we obtain the value of $E_{\text{C}} = 0.51$\,meV for $N_{\text{a}} =10^{11}$\,cm$^{-2}$. The magnitude of $E_{\text{C}}$ sets the temperature scale above which conductance quantization deteriorates and  hole mobility $\mu_h$ decreases for QWs at the topological phase transition. Thus, the Coulomb gap model explains the stability of the QSHE up to 100\,K in WTe$_2$ monolayers \cite{Wu:2018_S}, where -- as mentioned above -- $E_{\text{C}}$ attains 10\,meV \cite{Song:2018_NC}.

\subsection{Electron-hole spin exchange}
\label{sec:exchange}
Taking into account previous insight that flip-flop transitions conserving total spin of edge electrons and, thus, due to spin-momentum locking edge current \cite{Tanaka:2011_PRL},   we are interested in determining the degree $r$ of axial symmetry breaking by the QW edge for the exchange interaction between topological edge electrons and acceptor holes
in HgTe QWs. One of possible mechanisms could be kinetic exchange discussed in semiconductors in the context of $p$--$d$ coupling \cite{Kacman:2001_SST} and the Kondo effect in quantum dots \cite{Pustilnik:2004_JPC}.
However, we demonstrate here that  the Bir--Pikus theory, originally developed  for excitons \cite{Bir:1974_B}, and later extended to
the case of exchange coupling between band electrons and acceptor holes in bulk semiconductors \cite{Sliwa:2008_PRB}, satisfactorily explains the experimental results for HgTe QWs.

The wave function of helical states at the edge along the $x$-direction assumes the form,
\begin{eqnarray}
\psi_e^{(+)}(\vec{r})& = &\frac{\exp(ikx)}{\sqrt{L_x}}[f_{1}(\vec{\mathtt{r}})u_1 + f_{4}(\vec{\mathtt{r}})u_4 + f_{3}(\vec{\mathtt{r}})u_3];\\
\psi_e^{(-)}(\vec{r})& = &{\mathcal{T}}\psi_e^{(+)}(\vec{r}),
\label{eq:general}
\end{eqnarray}
where $L_x$ is the edge length; $f_i$ are electron envelope functions that depend on
$\vec{\mathtt{r}} = (y,z)$;
$u_i$ are the relevant Luttinger-Kohn amplitudes, and
${\mathcal{T}} = -i\sigma_yK$ is a time reversal operator. Similarly, for the acceptor ground state,
\begin{eqnarray}
\psi_h^{(+)}(\vec{r}) &= &{\mathcal{C}}[F_{1h}(\vec{r})u_1 + F_{4h}(\vec{r})u_4];\\
\psi_h^{(-)}(\vec{r})& = &{\mathcal{T}}\psi_h^{(+)}(\vec{r}),
\end{eqnarray}
where ${\mathcal{C}} = \sigma_yK$ is a charge conjugation operator transforming the acceptor wave function from the
electron representation, employed in previous sections for the determination of electronic states,
to the hole representation relevant here. In contrast, in the case of edge states we consider electrons and, thus,
 the electron representation for the Fermi level both above and below the Dirac point.

With these wave functions we determine a long-range contribution to the electron-hole exchange \cite{Bir:1974_B,Sliwa:2008_PRB}. This interaction is nonlocal and
represented, in the case under consideration, by the matrix elements,
%\begin{widetext}
\begin{eqnarray}
{\cal{H}}_{eh}^{\sigma,\sigma\prime;\mu,\mu\prime} = -\frac{\hbar^2}{m_0^2}\sum_{\alpha,\beta}\int d\vec{r}_1d\vec{r}_2\partial^2_{r_{1}^{(\alpha)}r_{2}^{(\beta)}}V(r_{12}) \nonumber\\
\cdot\langle\psi_e^{\sigma}(\vec{r}_1)|p_{\alpha}|\psi_h^{\mu}(\vec(r_1)\rangle_{\Omega} \langle\psi_e^{\sigma\prime}(\vec{r}_2)|p_{\beta}|\psi_h^{\mu\prime}(\vec(r_2)\rangle_{\Omega},
\label{eq:Heh}
\end{eqnarray}
%\end{widetext}
where $\sigma$ and $\mu = \pm$; $\alpha,\beta = x,y,z$; ${\vec{r}}_{12} = \vec{r}_1 - \vec{r}_2$. Furthermore, in accord with the effective mass theory, internal matrix elements containing momentum operators $p_{\alpha}$ and $p_{\beta}$
are over the elementary cell volume $\Omega$ and involve the amplitudes $u_i$, whereas the envelope functions are assumed constant within this volume. As seen, non-vanishing matrix elements correspond to coupling of $s$-type and $p$-type edge states with $p$-type and $s$-type acceptor states, respectively. Neglecting electrostatic image charges,
the Coulomb energy has a standard form,
\begin{equation}
V(r_{12}) = \frac{e^2}{4\pi\epsilon_0\epsilon r_{12}}.
\end{equation}
It is convenient to present a spatial function in the Coulomb term as a sum of the local monopole and non-local dipole components,
\begin{eqnarray}
\partial^2_{r_{1}^{(\alpha)}r_{1}^{(\beta)}}r_{12}^{-1} = -\frac{4\pi}{3}\delta_{\alpha \beta}\delta(\vec{r}_{12}) \nonumber \\
- \frac{(r_{12})^2\delta_{\alpha \beta} -3\vec{r}_{12}^{(\alpha)}\vec{r}_{12}^{(\beta)}}{r_{12}^5}.
\label{eq:monopole_dipole}
\end{eqnarray}

In the case of electrons at the bottom of the conduction band and holes localized on acceptor impurities in bulk zinc-blende semiconductors, the exchange coupling has a scalar (Heisenberg) form, ${\cal{H}}_{eh} = -{\cal{J}}\vec{s}\cdot\vec{j}$, where $j = 3/2$ and ${\cal{J}}< 0$ \cite{Sliwa:2008_PRB}. Since the QW and the edge break rotational symmetry,
we expect, in the presence of intratomic spin-orbit coupling, a nonscalar form of the exchange interaction,
\begin{equation}
{\cal{H}}_{eh} = -\sum_{\alpha,\beta}s_{\alpha}{\cal{J}}^{(\alpha,\beta)}j_{\beta},
\label{eq:Heh_J}
\end{equation}
where ${\cal{J}}^{(\alpha,\beta)}$ is a real tensor and, in a standard notation, ${\cal{J}}_{\alpha} ={\cal{J}}^{(\alpha,\alpha)}$,
 and if $\alpha \ne \beta$, ${\cal{J}}_{\alpha,\beta}^{(s)} =({\cal{J}}^{(\alpha,\beta)}+ {\cal{J}}^{(\beta,\alpha)}/2$;
 $D_{\alpha} = \sum_{\beta,\gamma}\epsilon_{\alpha\beta\gamma}{\cal{J}}^{(\beta,\gamma)}/2$, where $D_{\alpha}$ are vector components
 of the Dzyaloshinskii-Moriya (DM) contribution and $\epsilon_{\alpha\beta\gamma}$ is the antisymmetric Levi-Civita tensor.

By comparing matrix elements of Hamiltonians given in Eqs.\,\ref{eq:Heh} and \ref{eq:Heh_J}, considering both monopole and dipole contributions
of the long-range electron-hole exchange interaction (Eq.\,\ref{eq:monopole_dipole}), and taking into account that ${\cal{H}}_{eh} = -{\cal{H}}_{ee}$, we arrive to
final forms of non-zero exchange tensor components and a spin-independent part of the Fock energy $\Delta_{eh}$,
\begin{widetext}
\begin{eqnarray}
{\cal{J}}_{x,y} = A\left(\frac{16\pi}{9}{\cal{M}}_{21} + \frac{4}{3}{\cal{M}}_{31}  -3{\cal{M}}_{52}  \pm {\cal{M}}_{51}\right);\\
{\cal{J}}_{z} =  A\left[\frac{8\pi}{9}({\cal{M}}_{11} -3{\cal{M}}_{12})  +\frac{5}{3}{\cal{M}}_{41} +  {\cal{M}}_{42}
                                              + 2\sqrt{3}i{\cal{M}}_{63}\right];\\
{\cal{J}}_{yz}^{(s)} = iA\left(\frac{8\sqrt{3}\pi}{9}{\cal{M}}_{13}  -  \frac{\sqrt{3}}{3}{\cal{M}}_{43}  -\sqrt{3}{\cal{M}}_{53}\right);\\
{\cal{D}}_{x} = -iA\left(\frac{8\sqrt{3}\pi}{9}{\cal{M}}_{13}  +  \frac{\sqrt{3}}{3}{\cal{M}}_{43}  + \sqrt{3}{\cal{M}}_{53}\right);\\
\Delta_{eh} = A\left[\frac{2\pi}{3}({-\cal{M}}_{11} + {\cal{M}}_{12}) - \frac{1}{4}{\cal{M}}_{41} - \frac{1}{4}{\cal{M}}_{42}
                 -\frac{\sqrt{3}i}{2}{\cal{M}}_{63}\right]. \label{eq:Delta}
\end{eqnarray}
\end{widetext}
In the above formulae the prefactor $A$ contains information about
the strength of the Coulomb interaction and $\vec{k}\cdot\vec{p}$ coupling,
\begin{equation}
A = \frac{e^2}{4\pi\epsilon_0\epsilon}\frac{P^2}{E_{eh}^2L_x},
\label{eq:A}
\end{equation}
where $P$ is Kane's $s$-$p$ momentum matrix element and $E_{he}$ is the energy distance between
hole and electron states in question.

Particular matrix elements ${\cal{M}}_{ij}$ are given by
\begin{equation}
{\cal{M}}_{ij} = \int d\vec{r}_1d\vec{r}_2g_i(\vec{r}_1,\vec{r}_2)O_j(\vec{r}_1,\vec{r}_2),
\end{equation}
where
\begin{eqnarray}
g_1({\vec{r}_1},{\vec{r}_2})& = &\delta(\vec{r}_{12});\\
g_2({\vec{r}_1},{\vec{r}_2})& = &\cos[k(x_1 + x_2)]\delta(\vec{r}_{12});\\
g_3({\vec{r}_1},{\vec{r}_2})& = &\cos[k(x_1 + x_2)](r_{12}^2 - 3z_{12}^2)/r_{12}^5; \\
g_4({\vec{r}_1},{\vec{r}_2})& = &\cos[k(x_1 - x_2)](r_{12}^2 - 3z_{12}^2)/r_{12}^5;\\
g_5({\vec{r}_1},{\vec{r}_2})& = &\cos[k(x_1 + x_2)](x_{12}^2 - y_{12}^2)/r_{12}^5; \\
g_6({\vec{r}_1},{\vec{r}_2})& = &\sin[k(x_1 - x_2)]x_{12}z_{12}/r_{12}^5,
\end{eqnarray}
Overlap functions between edge-electron and acceptor-hole envelopes read
%\begin{widetext}
\begin{eqnarray}
O_1(\vec{r}_1,\vec{r}_2) = [f_4(\vec{\mathtt{r}}_1)F_{1h}(\vec{r}_1) - f_1(\vec{\mathtt{r}}_1)F_{4h}(\vec{r}_1)] \nonumber \\
                              \cdot[f_4(\vec{\mathtt{r}}_2)F_{1h}(\vec{r}_2) - f_1(\vec{\mathtt{r}}_2)F_{4h}(\vec{r}_2)];\\
O_2(\vec{r}_1,\vec{r}_2) = f_3(\vec{\mathtt{r}}_1)F_{1h}(\vec{r}_1)f_3(\vec{\mathtt{r}}_2)F_{1h}(\vec{r}_2);\\
O_3(\vec{r}_1,\vec{r}_2) = f_3(\vec{\mathtt{r}}_1)F_{1h}(\vec{r}_1) \nonumber \\
                         \cdot[f_4(\vec{\mathtt{r}}_2)F_{1h}(\vec{r}_2) - f_1(\vec{\mathtt{r}}_2)F_{4h}(\vec{r}_2)],
\end{eqnarray}
where for the adopted phase convention, in accord with the results presented in Fig.~\ref{fig:envelope}, $f_1(\vec{\mathtt{r}}), f_3(\vec{\mathtt{r}})$, and $F_{1h}(\vec{r})$ are real, $f_3(\vec{\mathtt{r}})$ is negative, whereas $f_4(\vec{\mathtt{r}})$ and $F_{4h}(\vec{r})$ are imaginary and change sign as a function of $z$.
%\end{widetext}

Inspection of the above equations shows that exchange integrals ${\cal{J}}_{\alpha}$ are negative, implying antiferromagnetic coupling between edge electrons and localized holes, which allows for the Kondo coupling. At the same time, the presence of terms breaking the axial symmetry leads to spin non-conserving transitions ($r\ne 0$) and, hence to net backscattering of edge electrons. The off-diagonal exchange tensor components ${\cal{J}}_{yz}^{(s)}$ and ${\cal{D}}_{x}$ are non-zero for both monopole and dipole coupling but only if the inversion symmetry is broken, i.e., the acceptor resides away of the QW center, so that the hole envelope functions  $F_{1h}$ and $F_{4h}(\vec{r})$ cease to be symmetric and antisymmetric with respect to $z = 0$, respectively. The presence of such terms was noted for dipole interactions of edge electrons with nuclear spins \cite{Lunde:2013_PRB} and for the Heisenberg interaction with magnetic impurities \cite{Kimme:2016_PRB}.
However, in our case, even if inversion symmetry is maintained and, moreover, even if the $s$-type component in the hole wave function is negligible ($F_{1h} = 0$), spin non-conserving transitions are still allowed by the edge-induced breaking of the axial symmetry, leading to ${\cal{J}}_x \ne {\cal{J}}_y$ in the dipole contribution.

%We are interested in the magnitude of $\Delta_{eh}$ and in the ratio $r$ of the rate corresponding to spin non-conserving transitions (determined by squares of ${\cal{J}}_x-{\cal{J}}_y$, ${\cal{J}}_{yz}^{(s)}$ and ${\cal{D}}_{x}$ to the rate of spin-flop processes (determined by square of ${\cal{J}}_x+{\cal{J}}_y$).
For numerical evaluations, guided by theoretical results obtained for the edge states \cite{Lunde:2013_PRB,Papaj:2016_PRB,Krishtopenko:2018_PRB} and our data presented in Secs.\,\ref{sec:bands} and \ref{sec:acceptors}, we assume the electron and hole envelope functions in an approximate form,
\begin{widetext}
\begin{eqnarray}
f_1(y,z) = 2e_s\exp(-y/b)\Theta(y)\cos(\pi z/d_{\text{QW}})\Theta(d_{\text{QW}}/2-|z|)/(bd_{\text{QW}})^{1/2};\\
f_4(y,z) = 2ie_p\exp(-y/b)\Theta(y)\sin(2\pi z/d_{\text{QW}})\Theta(d_{\text{QW}}/2-|z|)/(bd_{\text{QW}})^{1/2};\\
f_3(y,z) = 2e_H\exp(-y/b)\Theta(y)\cos(\pi z/d_{\text{QW}})\Theta(d_{\text{QW}}/2-|z|)/(bd_{\text{QW}})^{1/2};\\
F_{1h}({\vec{r}}) = h_s\exp[-(x^2 +(y-y_{\text{m}})^2)^{1/2}/a^*]\Theta(y)\cos[\pi (z-z_{\text{m}})/d_{\text{QW}}]\Theta(d_{\text{QW}}/2-|z|)/N_{\text{o}};\\
F_{4h}({\vec{r}}) = ih_p\exp[-(x^2 +(y-y_{\text{m}})^2)^{1/2}/a^*]\Theta(y)\sin[2\pi (z-z_{\text{m}})/d_{\text{QW}}]\Theta(d_{\text{QW}}/2-|z|)/N_{\text{o}}.
\label{eq:simple}
\end{eqnarray}
\end{widetext}
Here $b$ describes the penetration length of the edge electron wave function into the QW; $\Theta(x)$ is the Heaviside step function,  $y_{\text{m}}$ and $z_{\text{m}}$ are distances of a maximum of the hole wave function from the sample edge located at $y =0$ and the QW center residing at $z=0$, respectively, and
$N_{\text{o}}$ is an in-plane normalization factor. We note that one expects $|z_{\text{m}}| < d_{\text{QW}}/2$ even for acceptors localized outside the QW, $|z_0|> d{\text{QW}}/2$. Making use of previous results  \cite{Lunde:2013_PRB}
as well as of our data presented in Secs.~\ref{sec:bands} and \ref{sec:acceptors}, we take $e_s = 0.26$; $e_p = 0.24$; $e_H = -0.93$; $h_s = 0.31$; and $h_p = 0.92$; as the factors determining the participation of particular orbital components in the total electron and hole wave functions. Furthermore, the computations have been performed for $d_{\text{QW}} = 8$\,nm and $a^* = 5$\,nm. A Monte Carlo method has been used to evaluate six dimensional integrals. Such a method minimizes systematic errors but necessarily leads to a statistical scatter of the results. The chosen number of evaluation points insures that the accuracy of the obtained data is better than one sigma.

\begin{figure}[tb]
	%\centering
%\hspace*{-2.3cm}
	\includegraphics[width =1.0\columnwidth]{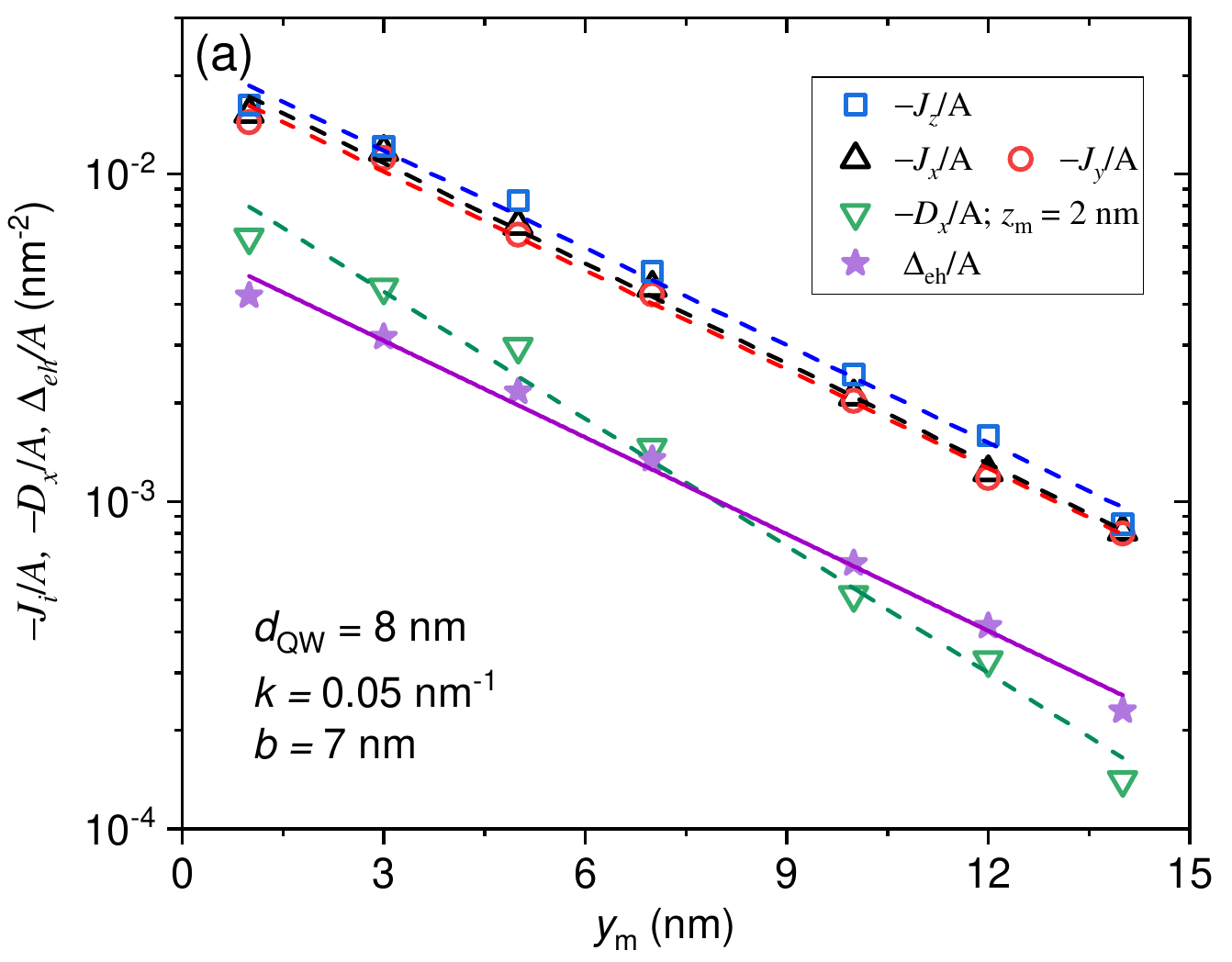}
   \includegraphics[width =1.0\columnwidth,angle=0]{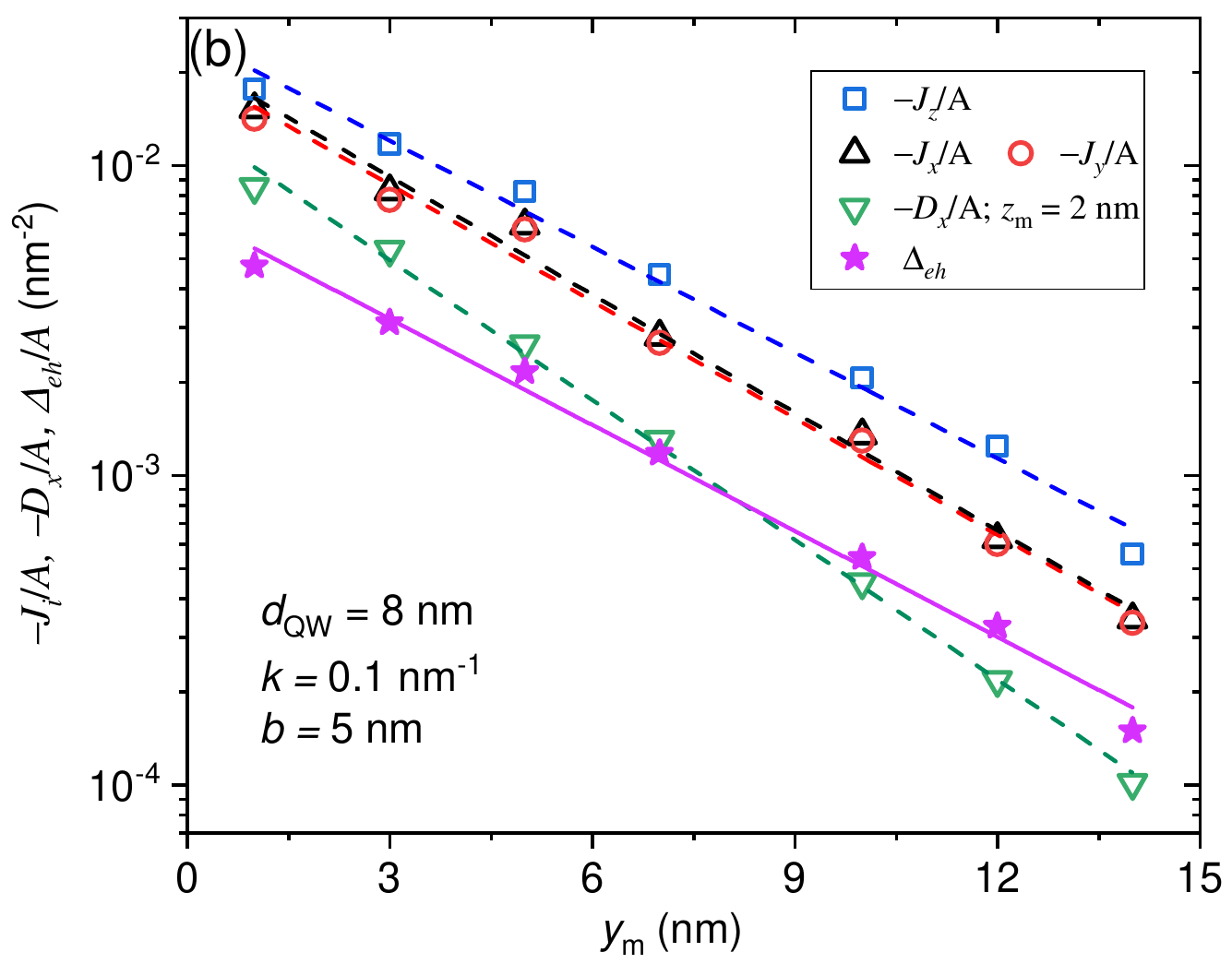}
	\caption{Computed values (points) of exchange integrals ${\cal{J}}_{\alpha}/A$,  the Dzyaloshinskii-Moriya energy $D_{x}/A$, and the spin-independent term $\Delta_{eh}$  for coupling between edge electrons and acceptor holes in topological HgTe QW, plotted as a function of the distance $y_{\text{m}}$ between the acceptor wave function maximum and the edge; (a) the electron wavevector $k = 0.05$\,nm$^{-1}$ and the penetration of helical state into the QW, $b =7$\,nm; (b) $k = 0.1$\,nm$^{-1}$ $b =5$\,nm. The material constant $A$ is defined in Eq.\,\ref{eq:A}. Lines are linear fits to the computed points.}
\label{fig:J}
\end{figure}

\begin{figure}[tb]
	%\centering
%\hspace*{-2.3cm}
	\includegraphics[width =\columnwidth, angle=0]{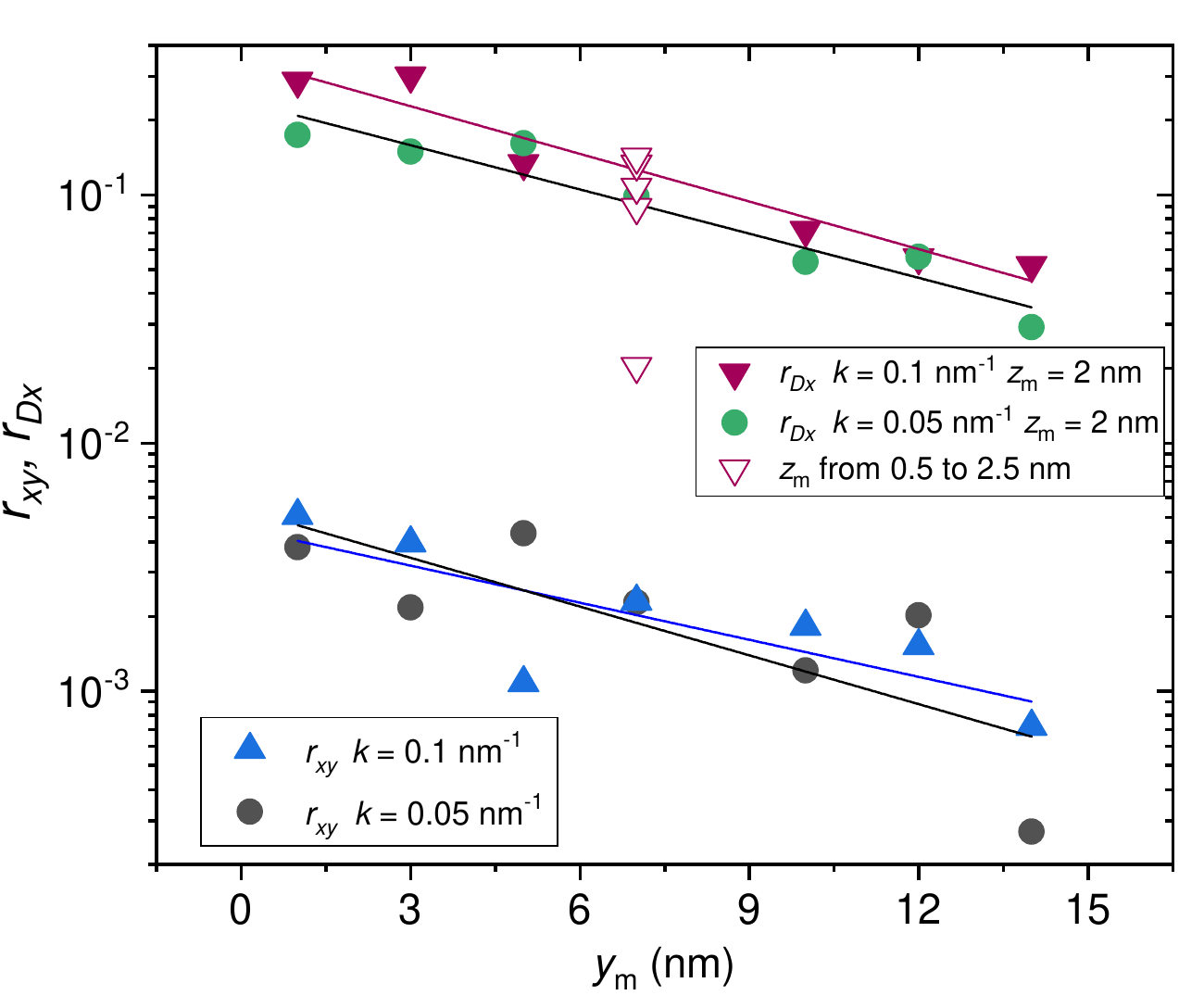}
	\caption{Computed magnitudes of
exchange anisotropy ratio  $r_{xy} = [(2({\cal{J}}_x-{\cal{J}}_y)/({\cal{{\cal{J}}}}_x+{\cal{J}}_y)]^2$ and $r_{Dx} = [(3D_x/({\cal{J}}_x+{\cal{J}}_y +{\cal{J}}_z)]^2$ for edge electrons in topological HgTe QWs as a function of the distance $y_{\text{m}}$ between the acceptor wave function maximum and the edge, obtained for ${\cal{J}}_{\alpha}$ and $D_x$ values display in Fig.~\ref{fig:J};  the results are for two electron wavevectors $k = 0.05$ and $0.1$\,nm$^{-1}$, for  which the penetration of helical state into the QW $b =7$ and $5$\,nm, respectively. In the case of $r_{Dx}$, the distance between  the QW center and the wave function maximum $z_{\text{m}} = 2$\,nm (full points) or varies between 0.5 and 2.5 nm with 0.5\,nm step (empty triangles); $k = 0.1$\,nm$^{-1}$; $r_{Dx}(z_{\text{m}} = 0) = 0)$. Lines are linear fits to the computed points.}
\label{fig:r}
\end{figure}

Figure \ref{fig:J} presents exchange tensor components as a function of $y_{\text{m}}$ for two values of $k = 0.05$ and 0.1\,nm$^{-1}$ and the corresponding magnitudes of $b = 7$ and 5\,nm, respectively \cite{Lunde:2013_PRB,Papaj:2016_PRB,Krishtopenko:2018_PRB}.
As could be expected for the exchange interaction, ${\cal{J}}_{\alpha}/A$ exponentially decay  with $y_{\text{m}}$ but, not surprisingly,  this decay is weaker for $r_{xy} = [(2({\cal{J}}_x-{\cal{J}}_y)/({\cal{J}}_x+{\cal{J}}_y)]^2$, as shown in Fig.~\ref{fig:r}. The magnitude of $r_{xy}$ is independent of $A$ and unaffected by a shift of the acceptor away from the QW center. However, for $z_{\text{m}}\ne 0$, other axial symmetry breaking terms appear, ${\cal{J}}_{yz}$ and $D_x$, where $|{\cal{J}}_{yz}|\ll |D_x|$, and,  for $z_{\text{m}} = d_{\text{QW}}/4$, $r_{Dx} = [(3D_x/({\cal{J}}_x+{\cal{J}}_y+{\cal{J}}_z)]^2 \gg r_{xy}$, as depicted in Fig.~\ref{fig:r}. In the subsequent two sections, the obtained values of ${\cal{J}}_{\alpha}/A$ and $r_i$ serve to estimate the magnitude of Kondo temperature and of the backscattering rate.

%For those parameters ${\cal{J}}_{x}/A = 0.124  0.1196$, ${\cal{J}}_{y}/A = 0.111  0.09972$, and ${\cal{J}}_{z}/A = 0.1079$.

Finally, we comment on the spin-independent part of the Fock energy $\Delta_{eh}$, which together
with Hartree terms originating from acceptor, edge, and gate charges, contribute to a self-consistent potential in the edge region, whose determination is beyond scope of the present work. Nevertheless, to have an idea about the energy scale involved,
we evaluate a contribution to one-electron energy resulting from the Fock term in a self-consistent way
making use of Eqs.\,\ref{eq:Delta} and \ref{eq:A}, and summing up over all holes with the areal density $N_h$,
\begin{equation}
E_{eh}^{(\text{F})} = \left[\frac{e^2N_{\text{a}}P^2}{4\pi\epsilon_0\epsilon}\int dy |f(y)|^2\Delta_{eh}(y)/A\right]^{1/3}.
\end{equation}
where the dependence $\Delta_{eh}(y)/A$ is displayed in Fig.~\ref{fig:J}  for two values of $k$ and $b$. Assuming $f(y) = \exp(-b/y)/\sqrt{b/2}$; $N_{\text{a}} = 0.5\cdot10^{11}$\,cm$^{-2}$; $P = 8.28\cdot10^{-8}$\,eVcm, and $\epsilon = 19.5$ we obtain $E_{eh}^{(\text{F})} = 4.8$ and $4.4$\,meV for $b = 5$ and 7\,nm, respectively.

\subsection{Kondo temperatures for acceptor holes and Mn spins}
\label{sec:Kondo}
\subsubsection{Acceptor holes}
A Kondo collective state results from antiferromagnetic exchange coupling of Fermi liquid with a single spin localized at ${\vec{R}}_i$. An antiferromagnetic electron-hole interaction leads also to the Kondo effect, as in the electron picture it corresponds to antiferromagnetic coupling of a surplus localized electron with the electron Fermi liquid.  To estimate an order of  magnitude of Kondo temperatures $T_{\text{K}}$,  we use a time-honored expression \cite{Daybell:1968_RMP} costed to the form,
\begin{equation}
k_{\text{B}}T_{\text{K}} \simeq W\exp[-1/(L^d\nu_d|{\mathcal{J}}_d|)],
\label{eq:TK}
\end{equation}
where $W$ is a energy width of carrier-containing states; $\nu_d$  is the carrier DOS per spin, and ${\cal{J}}_d$ represents the exchange energy of the antiferromagnetic interaction between spins of one carrier and a single localized paramagnetic center. Both $\nu_d$ and ${\mathcal{J}}_d$ correspond to values at the Fermi level, and depend on the dimensionality $d$ of the Fermi liquid residing in a structure of the volume $L^d$.
%In the case of electrons in helical topological states we expect a broad distribution of $T_{\text{K}}$ values, down to zero for paramagnetic spins localized away from the edges, $y_{\text{m}} \gg b$. %Accordingly we focus on obtaining an upper limit of the $T_{\text{K}}$ distribution.

For the case under consideration, $\nu_1 = 1/2\pi\hbar v_{\text{F}}$, where $v_{\text{F}} = 4\cdot10^5$\,m/s \cite{Lunde:2013_PRB,Papaj:2016_PRB,Krishtopenko:2018_PRB}, and ${\cal{J}}_1  = ({\cal{J}}_{x}+{\cal{J}}_{y})/2$ at given $y_{\text{m}}$, $k$, $b$, and the energy of the acceptor hole with respect to the Fermi level, $E_{eh}$.  Furthermore, together with $P$ and $\epsilon$  values quoted in previous sections, we take $W$ as a half of the bandgap, i.e., 15\,meV for $T_{\text{K}}$ estimations.

\begin{figure}[tb]
	%\centering
%\hspace*{-2.3cm}
	\includegraphics[width =1.0\columnwidth]{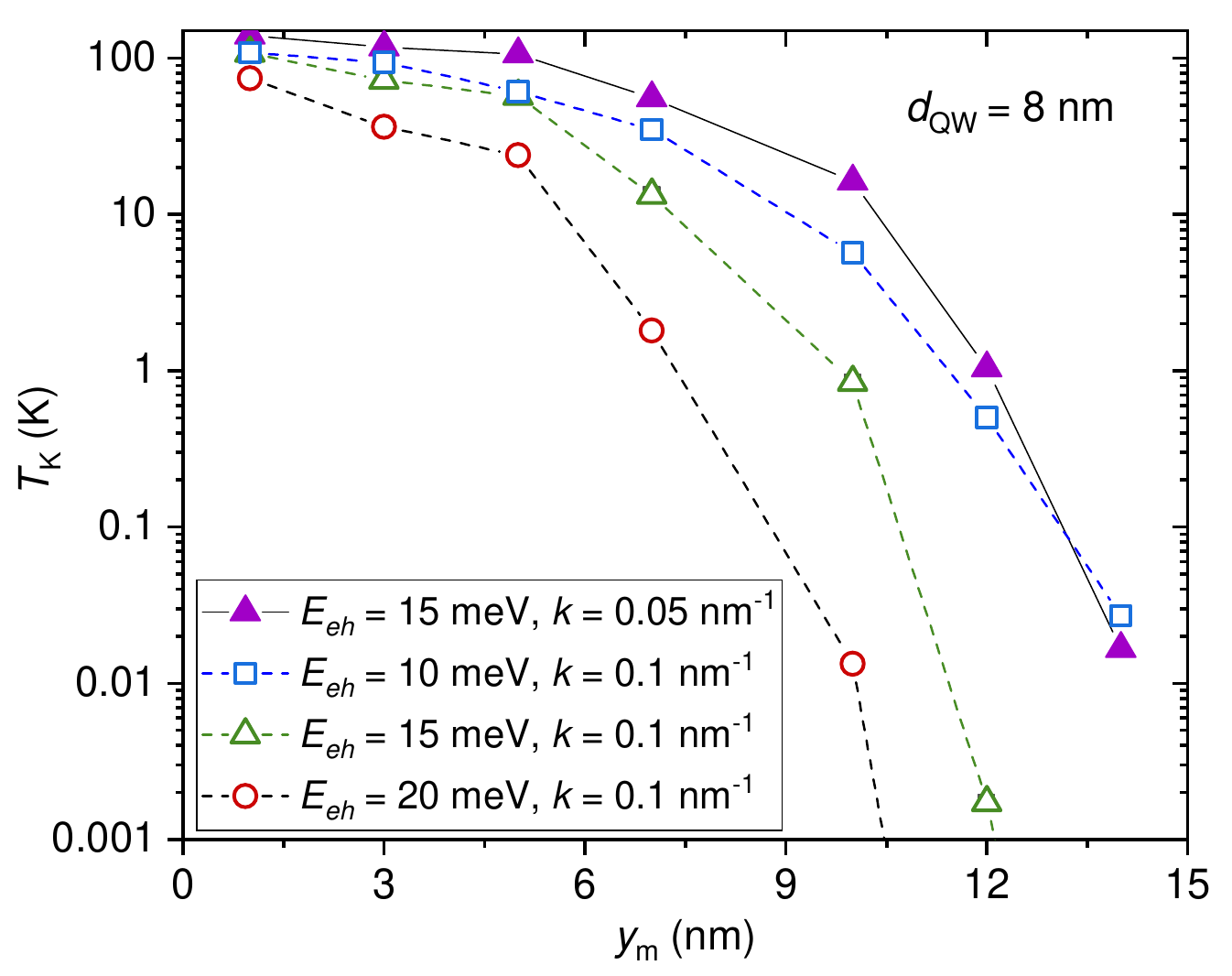}
	\caption{Calculated magnitudes (points) of Kondo temperatures, obtained for ${\cal{J}}_{x}$ and ${\cal{J}}_{y}$ values display in Fig.~\ref{fig:J}, for edge electrons in HgTe QWs containing holes bound to acceptors as a function of the distance $y_{\text{m}}$ between the acceptor wave function maximum and the edge calculated for various values of the electron-hole energy interval $E_{eh}$; full points: electron wavevector $k = 0.05$\,nm$^{-1}$ and the penetration of helical state into the QW, $b =7$\,nm; open points: $k = 0.1$\,nm$^{-1}$ $b =5$\,nm. Lines connect calculated points. }
\label{fig:TK}
\end{figure}

Several important conclusions emerge from $T_{\text{K}}$ magnitudes presented in Fig.~\ref{fig:TK} as a function of $y_{\text{m}}$. First, for acceptor holes localized close to the edge, $y_{\text{m}}< b$, $T_{\text{K}}$ reaches 100\,K, indicating that the Kondo physics is relevant for the system of edge electrons and acceptor holes, making competing effects, such as exchange interactions between hole spins, irrelevant. Second, due to the exponential dependencies of $T_{\text{K}}$ on ${\cal{J}}_{\alpha}$ and
${\cal{J}}_{\alpha}$ on $y_{\text{m}}$ there is a sharp cut-off $y{\text{c}}$, beyond which strong coupling of holes and electrons vanishes. The magnitude of $y{\text{c}}$ depends on $b$ and $E_{eh}$, that is, also on the position of the Fermi level in the acceptor band. Third, a broad distribution of $T_{\text{K}}$ values means that at any experimentally relevant temperature, coupling of edge electrons and holes corresponds to a superposition of weak and strong coupling limits, including the crossover between them occurring at $T \simeq T_{\text{K}}$.

\subsubsection{Kondo effect for Mn spins}
For comparison, we evaluate also the magnitude of Kondo temperature for a transition metal impurity, such Mn, residing at $\vec{R}_i$ in a HgTe bulk sample or in a topological edge channel. A $p$-type part of the Bloch function is relevant, as the $p$--$d$ exchange integral $\beta$ is antiferromagnetic, whereas the $s$--$d$ integral $\alpha$ is ferromagnetic \cite{Dietl:1994_HB}. Accordingly, ${\cal{J}}_d$ appearing in Eq.\,\ref{eq:TK} for $T_{\text{K}}$, can be written in a form,
\begin{equation}
{\cal{J}}_d = \beta [|F_{3/2}(\vec{R}_i)|^2 + |F_{1/2}(\vec{R}_i)|^2/3],
\end{equation}
where $F_{jz}$ are envelope functions accompanying the Kohn-Luttinger amplitudes $p_{3/2,j_z}$. We see that an upper limit
for the bulk case is ${\cal{J}}_3 = \beta/{\cal{V}}$, where ${\cal{V}}$ is the sample volume. Similarly, for a Mn ion localized at a topological channel, where $\vec{R}_i = (x,0,0)$ and, thus, ${\cal{J}}_1 = 4\beta/(L_xbd_{\text{QW}})$.

Now, knowing that $N_0\beta = -0.7$\,eV in HgTe  \cite{Autieri:2021_PRB}, where $N_0 = 1.49\cdot10^{22}$\,cm$^{-3}$ is the cation concentration, we are in position to evaluate a lower limit of the exponent $w$ in Eq.\,\ref{eq:TK} for $T_{\text{K}}$. For the bulk case,  where  $\nu_3 = k_{\text{F}}m^*/2\pi^2\hbar^2$ and $m^* =0.45m_0$  \cite{Jedrzejczak:1976_pssb}, $w = -1.1\cdot10^3$ for the hole concentration $p = 10^{16}$\,cm$^{-3}$ and $w = -49$ for  $p = 10^{20}$\,cm$^{-3}$, implying $T_{\text{K}} \sim 10^{-20}$\,K. For topological edge channels in HgTe QWs, taking $b = 5$\,nm, $d_{\text{QW}} =8$\,nm, and $v_{\text{F}} = 4\cdot10^5$\,m/s, we obtain $w = -350$.

Hence, it appears that for standard values of the $sp$-$df$ exchange integrals, no presence of the Kondo effect is expected for spins tightly localized  on $d$ or $f$ shells in semiconductors with magnetic ions. The same conclusion holds for nuclear spins for which the hyperfine coupling constant is at least four orders of magnitude smaller compared to $N_0\beta$.

%an upper limit of $T_{\trxt{K}$ is obtained for Mn ions residing close to the edge, where ${\cal{J}}_1 =|e_{p}F_e|^2\beta/3$, whose upper limit becomes ${\cal{J}}_1 = 4|e_{p}|^2\beta/3L_xbd_{\text{QW}}$, where $|e_{p}|^2$ is a fractional contribution of the Kohn-Luttinger amplitude $p_{3/2,\pm 1/2}$ in the edge electron wave function, taken here as 0.5;  $N_0\beta = -0.7$\,eV is the antiferromagnetic $p$--$d$ exchange energy \cite{Autieri:2021_PRB}, and $N_0 = 1.49\cdot10^{22}$\,cm$^{-3}$ is the cation concentration.
%
%Those parameters, together with $d_{\text{QW}} =8$\,nm and $b = 7$\,nm  imply that the magnitude of the exponent in the formula for $T_{\text{K}}$ (Eq.\,\ref{eq:TK}) becomes rather sizable,  $w = -2.0\cdot10^3$. Similarly, for heavy holes in bulk HgTe, where ${\cal{J}}_3 = \beta/V$ and $\nu_3 = k_{\cite{F}}m^*/2\pi^2\hbar^2$, where $m^* =0.45m_0$, \cite{Jedrzejczak:1976_pssb}, $w = -1.1\cdot10^3$
%for the hole concentration $p = 10^{16}$\,cm$^{-3}$ and $w = -49$ for  $p = 10^{20}$\,cm$^{-3}$.

\section{Comparison to experimental results}
\subsection{Topological protection length: acceptor holes and Mn spins}
\label{sec:protection}
\subsubsection{Acceptor holes}

Our approach to charge transport by topological edge channels in quantum spin Hall (QSH) materials is built on several pillars put previously forward  by others, discussed in the companion paper \cite{Dietl:2023_PRL}, or elaborated in previous sections of the present paper. First, charge dopants determine pertinent properties of 2D topological systems, including the dependence of carrier density on the gate voltage $V_{\text{g}}$ and the magnitudes of electron and hole mobilities. Second, owing to a dependence of the dopant binding energy on the position with respect to the QW center, the acceptors form a band extending over the whole bandgap. The associated Coulomb gap in the acceptor hole spectrum controls a contribution of QW states to charge transport in the QSH regime \cite{Dietl:2023_PRL}. Third, due to a close energetic proximity, exchange interactions between edge channel electrons and acceptor holes are strong enough to bring the system to the Kondo regime (Sec.~\ref{sec:Kondo}).  As known,  spin-dephasing rate reaches a unitary limit in that regime \cite{Maciejko:2009_PRL,Micklitz:2006_PRL}. Fourth, since the magnitude of Kondo temperature $T_{\text{K}}$ exponentially depends on the distance of the acceptor hole to the edge and on the energy interval to the Fermi level, $T_{\text{K}}$ values show a broad distribution covering the whole experimentally relevant temperature range (Sec.~\ref{sec:Kondo}). This effect determines also a spacial region from which acceptor holes contribute to spin-dephasing. Fifth, it has been emphasized by many authors \cite{Tanaka:2011_PRL,Altshuler:2013_PRL,Lunde:2013_PRB,Eriksson:2013_PRB,Kimme:2016_PRB} that because of spin-momentum locking, spin-conserving transitions imply momentum conservation, i.e., no net backscattering between helical channels in 2D topological insulators. However, spin-orbit interactions can result in spin-nonconserving processes described by nonscalar terms in the exchange coupling between edge electrons and paramagnetic centers \cite{Tanaka:2011_PRL,Altshuler:2013_PRL,Lunde:2013_PRB,Eriksson:2013_PRB,Kimme:2016_PRB}. The magnitude of such terms has been determined in Sec.~\ref{sec:exchange} for the case of QW holes bound to acceptor impurities.

We are interested in the topological protection length $L_{\text{p}} = v_{\text{F}}/\gamma_{\text{b}}$, where $\gamma_{\text{b}}$ is a backscattering rate for helical states. To see its relation to the two terminal conductance $G = 1/R$, we note that $R$ is non-zero due to the quantum contact resistance $R_{\text{q}} =h/2e^2$ and  the backscattering term which, according to the Einstein relation, is given by  $R_{\text{b}} = L^{2-d}\gamma_{\text{b}}(2e^2\nu_dv_{\text{F}}^2/d)^{-1}$. Hence, in our case, where $d =1$,
\begin{equation}
G = (2e^2/h)(1+ L_x/L_{\text{p}})^{-1}.
\label{eq:G}
\end{equation}

To evaluate $\gamma_{\text{b}}$ we recall a form of the carrier spin-dephasing rate in Kondo systems,
\begin{equation}
\gamma_s =  Cn_Sf(T/T_{\text{K}})/2\pi\hbar\nu_d,
\end{equation}
where $C$ is a numerical coefficient close to one; $n_S$ in the concentration of magnetic impurities and $f(x)$ is a function obtained by the numerical renormalization group approach \cite{Micklitz:2006_PRL,Costi:2009_PRL} that is more accurate than the original Nagaoka-Suhl resummation result, $f(x) = a/[a +\ln^2(T/T_{\text{K}})]$, where $a = 3\pi^2/4$. In either case, $f(0) =0$; $f(1) \simeq 1$, and $f(10) \simeq 0.6$.

It has been previously noted that compared to $\gamma_s$,  the backscattering rate for helical channels is reduced by the exchange anisotropy ratio $r$, $\gamma_b = r\gamma_s$  \cite{Tanaka:2011_PRL,Lunde:2013_PRB,Kimme:2016_PRB}.  Thus,  inverse $L_{\text{p}}$ can be written in an appealingly simple form \cite{Dietl:2023_PRL},
\begin{equation}
 L_{\text{p}}^{-1}(T) = \sum_ir^{(i)}f(T/T_{\text{K}}^{(i)})/L_x,
 \label{eq:Lp}
\end{equation}
where the summation is over all QW holes bound to acceptors for a given $V_{\text{g}}$.  As shown in Fig.~\ref{fig:r}, the $r$ value is not universal, but varies with the hole position in respect to the edge and QW center, $y_{\text{m}}$ and $z_{\text{m}}$, respectively, and to a lesser degree with $k$ and $b$. Similarly, according to results presented in Fig.~\ref{fig:TK}, the $T_{\text{K}}$ magnitude strongly depends  on $y_{\text{m}}$ and also on $E_{eh}$ that is controlled by the Fermi level position and $z_{0}$, the distance of the acceptor to the QW center, as shown in Fig.\,1 of Ref.~\onlinecite{Dietl:2023_PRL}. However, just to see whether we are on the right track, we take the areal hole density as $N_h =0.5\cdot10^{11}$\,cm$^{-2}$, an average value of $r^{(i)}$ as $r_{Dx}(z_{\text{m}} = 2$\,nm) = 0.13 (see, Fig.~\ref{fig:r}), the cut-off length beyond which strong coupling of holes and electrons tends to vanish $y_{\text{c}} = 2b = 10$\,nm (see, Fig.~\ref{fig:TK}), and an average value of $f(T/T_{\text{K}}) = 0.4$. These numbers lead to the  linear density $n_h = 5/\mu$m of holes participating in backscattering and $L_{\text{p}} = 4$\,$\mu$m, the order of magnitude consistent with experimental findings \cite{Konig:2007_S,Lunczer:2019_PRL,Majewicz:2019_PhD}. A small number of scattering centers has several important consequences, as discussed below.

\begin{figure}[tb]
	%\centering
%\hspace*{-2.3cm}
	\includegraphics[width =\columnwidth]{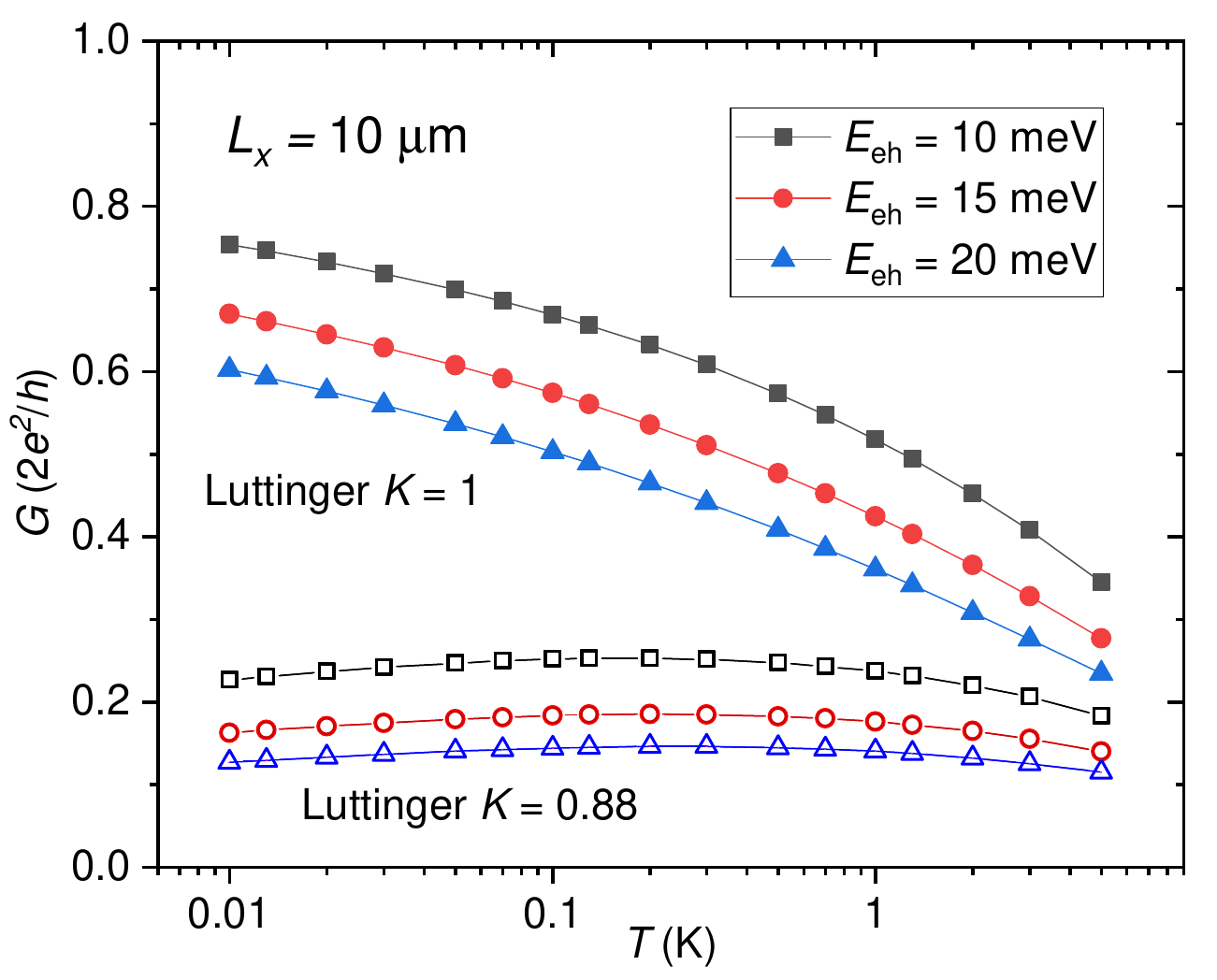}
	\caption{Configurationally averaged two-terminal conductance $G$ in the units $2e^2/h$ {\em{vs.}} temperature for a HgTe QW of the thickness $d_{\text{QW}} =8$\,nm and length $L_x = 10$\,$\mu$m computed for various energy distance of the Fermi level to the acceptor state $E_{eh}$; $k = 0.1$\,nm$^{-1}$ $b =5$\,nm; full points: no Luttinger liquid effects ($K = 1$); empty points: Luttinger liquid effects taken into account ($K = 0.88$). Lines connect calculated points.}
\label{fig:G}
\end{figure}

Figure~\ref{fig:G} shows the temperature dependence of two-terminal conductance $G$, as defined in Eq.~\ref{eq:G}, for a 10\,$\mu$m long device of HgTe QW. The values of $G$ have been obtained by integrating $L_{\text{p}}^{-1}$ given in Eq.~\ref{eq:Lp} in the $xy$ plane with the areal density of acceptor holes $N_h =0.5\cdot10^{11}$\,cm$^{-2}$. The latter changes with the gate voltage and its overall magnitude vary from sample to sample. However, typically, the gate voltage width corresponding to the gap region, $\Delta V_{\text{g}}$, implies the total acceptor density of the order of $10^{11}$\,cm$^{-2}$. The theoretical results are presented for three values of $E_{eh}$ and $L_x$. As $r(y)$, we have used the values fitted to $r_{Dx}(z_{\text{m}} = 2$\,nm) in Fig.~\ref{fig:r}. Similarly, $T_{\text{K}}(y)$ is determined from fitted values of ${\cal{J}}_x$ and ${\cal{J}}_y$ shown by dashed lines in Fig.~\ref{fig:J}.  Finally, the function $f(T/T_{\text{K}})$ has been obtained from the data in Fig.~3 for $d = 1$ of Ref.\,\onlinecite{Micklitz:2006_PRL} and from the proposed there high-temperature extrapolation.

A number of worthwhile conclusions can be drawn from data in Fig.~\ref{fig:G}. In particular, lowering of temperature eliminates
from backscattering spin centers with $T_{\text{K}} > T$, so that $L_{\text{p}}$ and $G$ are seen to steadily increase on cooling.
However, the predicted recovery of $G = 2e^2/h$ at $T \rightarrow  0$ \cite{Maciejko:2009_PRL,Vayrynen:2016_PRB} is not found,
as at any temperature there are centers far away from the channel for which $T_{\text{K}} \gtrsim T$.
Nevertheless our results are not in accord with experimental findings in a sense that no systematic increase
of $G$ on decreasing $T$ has been observed. This disagreement can point to the onset of localization,
predicted for disordered Kondo systems \cite{Altshuler:2013_PRL} or to the presence of Tomonaga-Luttinger effects
\cite{Maciejko:2009_PRL,Vayrynen:2016_PRB}. In has been shown \cite{Vayrynen:2016_PRB} that renormalization
group equations combining Kondo and carrier correlation phenomena imply a growth of the exchange
anisotropy ratio $r$ with lowering temperature down to $T^*$ according to,
\begin{equation}
r(T) = r(W)(W/k_{\text{B}}T)^2(1-K),
\end{equation}
with
\begin{equation}
k_{\text{B}}T^* = W\left[\frac{L_x\nu_1|{\cal{J}}_1|}{\sqrt{2}(1-K)}\right]^{1/(1-K)},
\end{equation}
where the Luttinger parameter $K < 1$ for interacting 1D systems. As shown in Fig.~\ref{fig:G}, incorporation of $r(T)$ into  our theory makes $G$ virtually independent of $T$ for a rather moderate interaction magnitude $K= 0.88$. It should be, however, noted that at any temperature only a part of acceptor holes resides sufficiently far from the edge to satisfy the condition $T> T^*$.

Up to now, we have discussed configuration-averaged behavior of an acceptors' containing system. However, a small number of acceptor holes involved in micron-size samples means that $G(V_{\text{g}})$ will show strong mesoscopic-like conductance fluctuations, as observed \cite{Konig:2007_S,Grabecki:2013_PRB,Shamim:2021_NC,Bubis:2021_PRB}. Furthermore, if at given $V_{\text{g}}$, acceptor holes close to the edge dominate
for which $T_{\text{K}} >T$, $f(T/T_{\text{K}})$ will increase and, hence, $G(T)$ decrease with temperature. By contrast, for distance acceptors $T_{\text{K}}<T$, so that a falling down of $f(T/T_{\text{K}})$ with $T$ works together with the Tomonaga-Luttinger effects to result in an increase of $G(T)$ on heating. Such changes in sign of $dG(T)/dT$ with  $V_{\text{g}}$ have been observed experimentally, though regions with $dG(T)/dT> 0$ appear to prevail \cite{Bubis:2021_PRB}. According to Eq.~\ref{eq:Lp} and many experimental observations, conductance quantization can be improved in short samples, $L_x \ll L_{\text{p}}$. However, a small number of relevant acceptors results in substantial conductance fluctuations. Charge traps biding two electrons in a singlet state might reduce backscattering and ensure pinning of the Fermi level in the QW band gap.

\subsubsection{Backscattering by Mn spins and precessional spin dephasing}
\label{sec:backscattering}
As there are typically $10^4$ more Mn spins with $S = 5/2$ in Hg$_{0.99}$Mn$_{0.01}$Te compared with the number of acceptor holes, a question arises about the role of Mn-induced backscattering of edge electrons. As the number of Mn ions in the edge region is also much larger than the number of edge electrons, we consider Mn spins in a continuous and classical approximation. Such an approximation is not only employed in the description of static and dynamic spintronics functionalities of ferromagnetic metals but has also been  found versatile and quantitatively accurate in the case of bound magnetic polarons and single quantum dot electrons immersed in a nuclear spin bath in semiconductors \cite{Dietl:2015_PRB} as well as when considering carrier-mediated ferromagnetism in DMSs \cite{Dietl:2014_RMP}.  Within this approach, it is convenient to expand local magnetization $\vec{M}(\vec{r})$ into a Fourier series,
\begin{equation}
\vec{M}(\vec{r}) = \sum_{\vec{q}}\vec{\eta}_{\vec{q}}\exp(i\vec{q}\cdot\vec{r})/\sqrt{{\cal{V}}}.
\end{equation}
For a uniform distribution of Mn ions over the volume ${\cal{V}}$ encompassing the QW and under thermal equilibrium conditions,
the fluctuation-dissipation theorem implies,
\begin{equation}
\langle\eta^{(\alpha)}_{\vec{q}}\eta^{(\beta)}_{\vec{q\prime}}\rangle = k_{\text{B}}T\chi_{\alpha,\beta}(\vec{q})\delta_{\vec{q},\vec{-q}\prime},
\end{equation}
where in cubic systems and in the absence of a magnetic field, the magnetic susceptibility tensor
$\chi_{\alpha,\beta}(\vec{q}) = \delta_{\alpha,\beta}\chi_{\text{Mn}}(q,T)$.
We then apply a standard weak-coupling procedure (Fermi's golden rule) for  spin-flip transitions between 1D helical states, $\psi_e^{(+)}(\vec{r}) \rightarrow \psi_e^{(-)}(\vec{r})$, and adopt the form of $\psi_e^{(\pm)}(\vec{r})$ given in Eq.~\ref{eq:general}. The scattering rate becomes,
%\begin{widetext}
\begin{equation}
\gamma_{\text{Mn}} = \frac{k_{\text{B}}T\chi_{\text{Mn}}(T)}{2\hbar^2g^2\mu_{\text{B}}^2v_{\text{F}}}
\int d\vec{\mathtt{r}}[\alpha^2|f_1(\vec{\mathtt{r}})|^4 +\beta^2(|f_3(\vec{\mathtt{r}})|^4 + |f_4(\vec{\mathtt{r}})|^4)],
\label{eq:gamma_Mn_1}
 \end{equation}
% \end{widetext}
where we have assumed that inverse $k$ and characteristic carrier confinement lengths are much longer than an average distance between magnetic ions, so that the $q$-dependence of magnetic susceptibility can be neglected, and $g = 2.0$ is the Land\'e factor of  Mn spins. For a paramagnetic case and using the form of the envelope functions given in Eq.~\ref{eq:simple},
\begin{equation}
\gamma_{\text{Mn}} = \frac{S(S+1)xN_0}{4\hbar^2v_{\text{F}}bd_{\text{QW}}}[\alpha^2|e_s|^4 + \beta^2(|e_H|^4 + |e_p|^4)].
 \label{eq:gamma_Mn_2}
 \end{equation}

For the $s$--$d$ exchange energy $N_0\alpha = 0.3$\,eV \cite{Autieri:2021_PRB},  $x = 0.01$ and other parameter values quoted in the previous paragraphs, the topological protection length becomes,
$L_{\text{p}} = 7.2$\,$\mu$m/$r_{\text{Mn}}$, where $r_{\text{Mn}} < 1$ is a relevant exchange anisotropy ratio.  In principle, $r_{\text{Mn}} \ll 1$ for  magnetic ions with the orbital momentum $L =0$. There are however mechanisms that can enlarge $r_{\text{Mn}}$. In particular,  Mn ions residing out of the QW center locally break  the inversion symmetry, which leads to the appearance of anisotropic exchange \cite{Kimme:2016_PRB}, similarly to the case of spin scattering by acceptor holes (Sec.~\ref{sec:exchange}). This mechanism should work at $x \rightarrow 0$.

However, beyond the limit $x \rightarrow 0$, interactions between magnetic ions is relevant. While the exchange mediated by edge electrons is weak \cite{Dietl:2014_RMP}, antiferromagnetic superexchange is significant. This interaction makes $\gamma_{\text{Mn}}$ to decrease with lowering temperature even stronger than $T\chi(T)$ implies because spin scattering becomes inelastic. In the spin-glass phase $T \le T_{\text{g}}(x)$, Mn spins cease to contribute to quantum decoherence and, accordingly, a recovery of universal conductance fluctuations was found in nanostructures of Hg$_{0.93}$Mn$_{0.07}$Te at $T< T_{\text{g}}$ \cite{Jaroszynski:1998_PRL}. All that might mean that Mn spins play a minor role in backscattering.

Actually, it was noted that the presence of random Rashba fields effectively enhances the $r$ value \cite{Kimme:2016_PRB}. We  argue that there are two other effects. First, as discussed in the next section, magnetic poloron formation around acceptor holes considerably weakness the Kondo effect and associated backscattering in Hg$_{1-x}$Mn$_{x}$Te compared to HgTe. Secondly, we suggest that dephasing of carrier spins by a dense bath of interacting magnetic moments  originates in semiconductors from a chain of spin precession events generated by local magnetization vectors rather than from flip-flop processes each involving a single magnetic ion.

In order to evaluate the precession-induced dephasing rate, we follow a time-honored Dykonov-Perel motional-narrowing approach to relaxation rates by spin-orbit fields, $\gamma_{\text{so}} \approx \tau_{\text{m}}\langle (g^*\mu_{\text{B}}B_{\text{so}}/2\hbar)^2/\rangle$, where, in the case under consideration, the motion time of the carrier wave packet extending over $\ell_x$,  $\tau_{\text{m}} = \ell_x/v_{\text{F}}$ with $k^{-1} \gg \ell_x \gg (N_0x)^{-1/3}$. This approach assumes that dynamics of Mn spins is slow, $\tau_{\text{m}} (\ell_x) < \tau_{\text{c}}(\ell_x)$, where $\tau_{\text{c}}$ is the Mn correlation time.  Qualitatively, the magnitude of local magnetization seen by a moving edge carrier,
\begin{equation}
\langle M_{\alpha}M_{\beta}\rangle =\delta_{\alpha\beta} k_{\text{B}}T\chi_{\text{Mn}}(T)\int d\vec{r}|\psi(\vec{r})|^4.
\end{equation}
For the wavepacket,
\begin{eqnarray}
\psi(\vec{r}) &=&  \frac{2\exp(ikx -x^2/4\ell_x^2 -y/b)\Theta(y)}{(2\pi)^{1/4}(\ell bd_{\text{QW}})^{1/2}} \nonumber\\
&&\cdot\cos(\pi z/d_{\text{QW}}\Theta(|z|-d_{\text{QW}}/2),
\end{eqnarray}
and noting that spin splitting $\Delta = {\cal{J}}_{sp-d}M/g\mu_{\text{B}}$ we arrive to,
\begin{equation}
\gamma_{\text{Mn}} = \frac{3S(S+1)xN_0{\cal{J}}_{sp-d}^2}{8\sqrt{\pi}\hbar^2v_{\text{F}}bd_{\text{QW}}}.
\label{eq:narrowing}
\end{equation}
We see that within a numerical factor of the order of one, the precession approach leads to the same expression, as given in Eqs.~\ref{eq:gamma_Mn_1} and \ref{eq:gamma_Mn_2}.
However, within such a model, similarly to the case of spin-transfer torque and electron precession around nuclear
spins in a quantum dot, a  change of carriers' spin momentum associated with backscattering is absorbed by the ensemble of Mn spins, so that  $r$ approaches 1. We conclude that backscattering by Mn spins may not be negligible in Hg$_{1-x}$Mn$_{x}$Te QWs, even in the absence of Rashba fields.

Finally, we return to the role of spin dynamics. The magnitude of $\tau_{\text{c}}$ is controlled by Mn spin diffusion and relaxation, determined by  scalar and  non-scalar terms in the Mn-Mn interaction Hamiltonian, respectively \cite{Dietl:1995_PRL}. We note that the mechanisms accounting for finite $\tau_{\text{c}}$ weakens building up of Mn magnetization by electric current. Furthermore, it was suggested that spin dynamics could promote depining of edge carriers localized by Kondo impurities \cite{Altshuler:2013_PRL}.

\subsection{Magnetic polaron gap and zero-field spin-splitting}
\label{sec:polaron}
We consider again isoelectronic magnetic impurities, such as Mn in II-VI compounds, in the paramagnetic phase. The presence of $sp$-$d$ exchange interactions affect, by the bound magnetic polaron (BMP) effect, donor electrons of acceptor holes even in the absence of macroscopic magnetization. Optical studies provided the evidence for the
presence of the acceptor BMP in Hg$_{1-x}$Mn$_x$Te \cite{Choi:1990_PRB,Zhu_JAP:2015}. A contribution of the BMP to thermally activated band conductivity \cite{Jaroszynski:1983_Pr} and to the Coulomb gap in the hoping region \cite{Terry:1992_PRL} was found in Cd$_{1-x}$Mn$_x$Te.

According to the analytical solution of the central spin problem, the polaron energy $\epsilon_{\text{p}}$ determines BMP energetics and thermodynamics  in the absence of an external magnetic field \cite{Dietl:1982_PRL},
\begin{equation}
\epsilon_{\text{p}}(T) =
{\cal{J}}_{sp-d}^2\chi_{\text{Mn}}(T)/[32(g\mu_{\text{B}}N_0)^2\pi a_{\text{B}}^3],
\end{equation}
where the magnitude of the exchange energy ${\cal{J}}_{sp-d}$ is given here by the values of the $s$--$d$ and $p$--$d$ exchange integrals $N_0\alpha$ and $N_0j_z\beta/3$, respectively,
weighted by the corresponding orbital content of the acceptor wave function;
$\chi_{\text{Mn}}(T)$ is the Mn susceptibility in the absence of acceptors;  $a_{\text{B}}$ is the acceptor localization radius determined from the participation number. In the case of the doubly-occupied acceptor with $J = 1$, $\epsilon_{\text{p}}(T)$ would be four times greater.

We identify the polaron gap between occupied and non-occupied acceptor centers, $E_{\text{P}}$, as the twice Fermi energy shift associated with the polaronic effect. For the doubly-degenerate acceptor state $E_{\text{P}}$ assumes the form \cite{Jaroszynski:1983_Pr},
\begin{equation}
E_{\text{P}} = \epsilon_{\text{p}} + 2k_{\text{B}}T\ln(1 + \epsilon_{\text{p}}/k_{\text{B}}T).
\end{equation}
For the $E_{3/2}$ state, ${\cal{J}}_{sp-d} = -0.7$\,eV implying, neglecting antiferromagnetic interactions between Mn spins, $E_{\text{P}} = 1.7$\,meV
at $T = 2$\,K, $x =0.024$, and $a_{\text{B}} = 5$\,nm. Similarly, for the more relevant $E_{1/2}$ case,
${\cal{J}}_{sp-d} = -0.24$\,eV, for which $E_{\text{P}} = 0.31$\,meV. For the $x$ value in question there are about 500 Mn ions within the volume visited by the acceptor hole. We conclude that
the formation of BMPs in Hg$_{1-x}$Mn$_x$Te may substantially enhance the magnitude of the gap at the Fermi level in the acceptor band compared to the case of HgTe.

We are also interested in the magnitude of acceptor hole spin-splitting $\Delta$. If larger than $k_{\text{B}}T_{\text{K}}$, BMPs reduce
spin dephasing in the Kondo regime and, thus, backscattering of edge electrons, which improves the precision of resistance quantization in the quantum spin Hall effect regime.
The most probable magnitude of zero-field splitting $\bar{\Delta}$ is given by an implicit equation \cite{Dietl:1982_PRL},
\begin{eqnarray}
\bar{\Delta}^2 - 2\bar{\Delta}\epsilon_{\text{p}}(T)\tanh(\bar{\Delta}/2k_{\text{B}}T) - 8\epsilon_{\text{p}}(T)k_{\text{B}}T=0.
%\bar{\Delta} = 2\epsilon_{\text{p}}(T) + 4k_{\text{B}}T \,\,\, {\mbox{for}}\,\, \epsilon_{\text{p}}(T) \gg k_{\text{B}}T;\\
%\bar{\Delta} = 2[2\epsilon_{\text{p}}(T)k_{\text{B}}T]^{1/2} + 8^{1/2}\epsilon_{\text{p}}(T) \,\,\, {\mbox{for}} \epsilon_{\text{p}}(T) \ll k_{\text{B}}T.
\end{eqnarray}
For the parameters quoted above we obtain $\bar{\Delta} =2.7$\,meV  and 0.54\,meV at 2\,K for the $E_{3/2}$ and $E_{1/2}$ level, respectively.

In the particular case of the sample with $x = 0.012$ studied in Ref.\,\cite{Shamim:2021_NC} and for the level $E_{1/2}$, $\bar{\Delta} \geq k_BT$ at $T \leq 3.5$\,K.
It is, therefore, clear that the presence of Mn, $via$ the formation of BMPs and the associated diminishing of the role played by the Kondo effect, can significantly reduce backscattering of edge electrons, leading to the recovery of quantized resistance at low temperature in magnetically doped QWs, as observed in a  Hg$_{0.988}$Mn$_{0.012}$Te QW at 0.2\,K \cite{Shamim:2021_NC}.

\section{Conclusions}
The quantitative results presented here support the view that acceptor states play a crucial role in the physics of quantum
spin Hall effect in HgTe quantum wells and related systems. On the one hand, the ionization of acceptors accounts for a non-zero width of the quantized plateaus and, on the other, the strong Kondo coupling of edge electrons and acceptor holes leads to the unitary limit of the spin-flip scattering rate. A non-zero orbital momentum specific to $p$-type Kohn-Luttinger amplitudes together with breaking of axial and inversion symmetry by the edge and off center hole location  allows for flow of edge electron angular momentum to crystal orbital momentum and, thus, for efficient backscattering, despite spin-momentum locking. According to the present insight, lowering of temperature drives the topological edge electrons from the Fermi, to the Kondo, and finally to the Luttinger liquid. Interestingly, the formation of bound magnetic polarons in magnetically doped samples weakness Kondo scattering, which allows for a recovery of conductance quantization at low temperatures, as observed \cite{Shamim:2021_NC}.

As discussed in the companion paper \cite{Dietl:2023_PRL}, the acceptor band model qualitatively elucidates several surprising properties in the vicinity of the topological phase transition, where $E_{\text{g}} \simeq 0$ and, thus, acceptor form resonant states. In particular, gate-induced discharging of the acceptor states explains an unexpectedly slow  rise of the itinerant hole concentration with increasing negative gate voltage \cite{Shamim:2020_SA,Yahniuk:2021_arXiv} and unusually wide integer quantum Hall plateaus in the same gate region \cite{Konig:2007_S,Yahniuk:2019_QM,Shamim:2020_SA}.  Furthermore, the presence of resonant bound magnetic polarons in magnetically doped samples diminishes Kondo scattering of electrons by acceptor holes and makes the Coulomb gap harder. We argue that these effects account for the hole mobility as large as $0.9\cdot10^6$\,cm$^2$/Vs at hole density as low as $p= 3.5\cdot10^9$\,cm$^{-2}$ in a Hg$_{0.98}$Mn$_{0.02}$Te QW at 20\,mK \cite{Shamim:2020_SA}. In the same way, the model qualitatively explains $\mu = 20\cdot10^6$\,cm$^2$/Vs at the electron concentration $n = 2.5\cdot10^{13}$\,cm$^{-3}$ at 2\,K in a bulk Hg$_{0.94}$Mn$_{0.06}$Te  under hydrostatic pressure that made possible a fine tuning of the system to the topological phase transition \cite{Sawicki:1983_Pr}.

We have also examined backscattering of electrons in helical states by localized spins in DMSs. Our results indicate that precessional spin dephasing by magnetization spacial fluctuations leads to sizable backscattering even in the spin-momentum locking case, as the electron spin momentum is transferred to the magnetic subsystem rather than to an individual magnetic ion.

\section*{Acknowledgments}
This work was supported  by the Foundation for Polish Science through the International Research
Agendas program co-financed by the European Union within the Smart Growth Operational Programme.

%merlin.mbs apsrev4-1.bst 2010-07-25 4.21a (PWD, AO, DPC) hacked
%Control: key (0)
%Control: author (0) dotless jnrlst
%Control: editor formatted (1) identically to author
%Control: production of article title (0) allowed
%Control: page (1) range
%Control: year (0) verbatim
%Control: production of eprint (0) enabled
%

%\bibliography{TI_td13Nov2022}	
\end{document}